\documentclass[fleqn,usenatbib]{mnras}

\usepackage{newtxtext,newtxmath}

\usepackage[T1]{fontenc}
\usepackage{ae,aecompl}

\usepackage{graphicx}	
\usepackage{amsmath}	
\usepackage[xindy]{glossaries}
\glsdisablehyper
\usepackage{caption}
\usepackage{subcaption}
\usepackage{siunitx}
\usepackage{tabularx}
\usepackage[dvipsnames]{xcolor}
\usepackage{hyperref}
\usepackage[normalem]{ulem}  

\usepackage{float} 

\newacronym{dm}{DM}{Dispersion Measure}
\newacronym{frb}{FRB}{Fast Radio Burst}
\newacronym{fwhm}{FWHM}{Full-Width at Half-Maximum}
\newacronym{ism}{ISM}{Interstellar Medium}
\newacronym{rfi}{RFI}{Radio-frequency Interference}
\newacronym{rm}{RM}{Rotation Measure}
\newacronym{ska}{SKA}{Square Kilometre Array}
\newacronym{sefd}{SEFD}{System Equivalent Flux Density}
\newacronym{snr}{S/N}{Signal-to-Noise Ratio}
\newacronym{tab}{TAB}{Tied-Array Beam}
\newacronym{pa}{PA}{position angle}
\newacronym{gp}{GP}{Giant pulse}
\newacronym{gps}{GPs}{Giant pulses}
\newacronym{rms}{RMS}{root mean square}
\newacronym{lmc}{LMC}{Large Magellanic Cloud}

\title[Giant Pulses of PSR~J0540$-$6919]{The Thousand-Pulsar-Array programme on MeerKAT III:\\
Giant pulse characteristics of PSR~J0540$-$6919}

\author[M. Geyer et al.]{
M.~Geyer,$^{1}$\thanks{E-mail: mgeyer@ska.ac.za},
M.~Serylak$^{1}$,
F. Abbate$^{2}$,
M.~Bailes$^{3,4}$,
S.Buchner$^{1}$,
J.Chilufya$^{5}$,
S.~Johnston$^{6}$,\newauthor
A.~Karastergiou$^{7}$,
R.~Main$^{2}$,
W.~van Straten$^{8}$,
M. Shamohammadi$^{3,4}.$
\\
$^{1}$ South African Radio Astronomy Observatory, 2 Fir Street, Black River Park, Observatory 7925, South Africa\\
$^{2}$ Max Planck Institut f\"ur Radioastronomie, Auf dem H\"ugel 69, 53121 Bonn, Germany\\
$^{3}$ Centre for Astrophysics and Supercomputing, Swinburne University of Technology, PO Box 218, Hawthorn, VIC 3122, Australia\\
$^{4}$ ARC Centre of Excellence for Gravitational Wave Discovery (OzGrav), Mail H29, Swinburne University of Technology, PO Box 218, Hawthorn, \\ VIC 3122, Australia \\ 
$^{5}$ University of Western Cape, Physics Department, Cape Town 7535, South Africa\\
$^{6}$ CSIRO Astronomy and Space Science, Australia Telescope National Facility, PO~Box~76, Epping NSW~1710, Australia\\
$^{7}$University of Oxford, Sub-Department of Astrophysics, Denys Wilkinson Building, Keble Road, Oxford, OX1 3RH, United Kingdom\\
$^{8}$Institute for Radio Astronomy \& Space Research,
Auckland University of Technology, Private Bag 92006, Auckland 1142, New Zealand \\
}

\date{Accepted 2021 May 14. Received 2021 April 09; in original form 2020 July 30}
\pubyear{2020}

\begin{document}
\label{firstpage}
\pagerange{\pageref{firstpage}--\pageref{lastpage}}
\maketitle


\begin{abstract}
PSR~J0540--6919 is the second-most energetic radio pulsar known and resides in the Large
Magellanic Cloud. Like the Crab pulsar it is observed to emit giant radio pulses (GPs). 
We used the newly-commissioned
PTUSE instrument on the MeerKAT radio telescope to search for GPs across three observations.
In a total integration time of 5.7 hrs we detected 865 pulses above our 7$\sigma$ threshold.
With full polarisation information for a subset of the data, we estimated the Faraday rotation measure, $\rm{RM}=-245.8 \pm 1.0$ rad m$^{-2}$ toward the pulsar. The brightest of these pulses is $\sim$60\% linearly polarised but the pulse-to-pulse
variability in the polarisation fraction is significant. 
We find that the cumulative GP flux distribution follows a power law
distribution with index $-2.75 \pm 0.02$. Although the detected GPs make up only $\sim$10\% of the mean flux, their average pulse shape is indistinguishable from the integrated pulse profile, and we postulate that there is no underlying emission.
The pulses are scattered at L-band frequencies with the brightest pulse exhibiting 
a scattering time-scale of $\tau = 0.92 \pm 0.02$\,ms at 1.2~GHz.
We find several of the giants display very narrow-band \textit{flux knots} similar to those seen in many Fast Radio Bursts, which we assert cannot be due to scintillation or plasma lensing. The GP time-of-arrival distribution is found to be Poissonian on all but the shortest time-scales where we find four GPs in six rotations, which if GPs are statistically independent is expected to occur in only 1 of 7000 observations equivalent to our data.
\end{abstract}



\section{Introduction}
\label{sec:intro}
Observations of single pulses from radio pulsars provide a rich vein of phenomenology which has been mined by observers and theorists alike over the past 50 years. Giant pulses (GPs, often referred to as \textit{giants} in this work) are one such class of single pulse emission and are loosely defined as the emission of short duration, very bright pulses often occurring in narrow phase windows.

Most pulsars display a log-normal distribution of pulse energies \citep{Johnston2001,Cairns2003,Burke-Spolaor2012} making pulses with ten times the mean energy rare. As such, energy thresholds were previously used to differentiate between normal and very bright or `giant' pulses in a straight-forward way (e.g. \citealt{Heiles1970}). More specifically, however, the flux distribution of such giant pulses is found to be well described by a power law that extends to values many times the mean flux density of the pulsar \citep{lcu+95, Johnston2004, kbm+06}. We use this description of pulses whose energies follow a power law distribution to define giant pulses in this paper.

Only a handful of pulsars emit giant pulses by this criterion. The Crab pulsar was discovered in the radio via its giant pulses \citep{sr68} and these were subsequently determined to align in rotational phase with high-energy X-ray and $\gamma$-ray emission \citep{lcu+95}. The young Crab pulsar, with a spin-down age of $\sim$1.2\,kyr, remained the only pulsar known to emit giant pulses until the surprising discovery of giant pulse emission in the older millisecond pulsar PSR~B1937+21 (spin-down age > 200 Myr,  \citealt{cstt96}). Since then other millisecond pulsars have been found to emit giant pulses (e.g. \citealt{rj01,kbm+06,ovb+14}) but the only other young pulsar (spin-down age < 10\,kyr) with giant pulse emission is PSR~J0540$-$6919 (B1950-name: B0540$-$69) with a characteristic age of 1.7\,kyr \citep{Zhang2001}.

PSR~J0540$-$6919 is located in the \gls{lmc} at a distance of $49.7 \pm 1.1$\,kpc \citep{Walker2011} and, due to their similarities in spin-down age and spin period, is often seen as an extragalactic twin of the Crab pulsar. The pulsar is located in the centre of the supernova remnant SNR~0540$-$693. PSR~J0540$-$6919 was first discovered in the X-rays \citep{shh84} with a spin period of 50.8\,ms. Optical pulsations were seen by \citet{mp85} but it was not until a decade later that weak radio emission was marginally detected by \citet{mml+93}. A decade later again, high time resolution observations of the pulsar at a centre frequency of 1.4~GHz using the Parkes telescope revealed the presence of giant pulses \citep{jr03}.

The pulsar was subsequently observed for a total of 72 hours with the Parkes telescope in 2003 \citep{jrmz04}. During that time 141 giant pulses were detected with mean flux densities > 4.5\,mJy. The pulses appeared at two distinct pulse phases and were aligned in phase with the X-ray emission. Such alignment is observed in other GP emitting pulsars including PSRs~J1824-2452A, B1937+21 and J2018+4232 \citep{rj01,Cusumano2003, kbm+06}. The GP flux densities followed a power-law distribution. From this sample, the mean flux density of the GPs were measured to be 10.4\,mJy in the `leading' emitting phase window and 7.5\,mJy in the `trailing' phase window and the brightest pulse detected had a mean flux density of 64\,mJy. The authors found that the cumulative flux density followed a power law with index $-1.8$, and presented the first detection of the faint integrated profile with a flux density of only 24\,$\si{\micro}$Jy \citep{jrmz04}. They surmised that the giant pulse emission contributes only a small fraction of this mean flux density but were limited by their telescope sensitivity to only the brightest giants.

Giant pulses from a cosmological population of neutron stars might also explain the origin of Fast Radio Bursts \citep[FRBs;][]{lbm+07,tsb+13,cw16}.
The repeating FRB~121102 appears to show power-law statistics \citep{lab+17}, similar to that of giant pulses; furthermore; FRB emission often appears to be band-limited \citep{hss+19}, as has been observed in nanoburst giant pulses from the Crab pulsar \citep{he07}. Such spectral morphology is often observed in repeating FRBs \citep{Andersen2019,Fonseca2020}, but whether the cause is intrinsic to the emission mechanism or a propagation effect in the FRB's immediate environment is as yet unclear.

In this paper we present the discovery of high signal-to-noise ratio GPs from PSR~J0540$-$6919 that are highly band-limited in nature, and consider whether these are intrinsic to this young pulsar's giant pulse emission mechanism or a result of interstellar scintillation.  We also consider whether all the detected flux from this pulsar could be GP emission only. 

These questions are addressed using highly sensitive observations of PSR~J0540$-$6919 made with the MeerKAT telescope \citep{Jonas2016, Camilo2018},  which at 20-cm wavelengths has four times the gain (2.8 K Jy$^{-1}$), and three times the bandwidth (856 MHz), compared to the Parkes radio telescope observations made in 2003. 
With this enhanced bandwidth and sensitivity we can better explore the broad-band nature, the flux density distribution and polarization characteristics of the giant pulse emission.

The paper proceeds as follows: Section \ref{sec:obs} describes the MeerKAT data, the reduction and calibration processes and how we search for GPs; Section \ref{sec:results} presents our results, including the integrated profile, the GP flux distribution and the characteristics of the discovered GP population. We end with discussions and conclusions in Section \ref{sec:disc} and \ref{sec:conclusion}. This work forms part of the Meertime Thousand-Pulsar-Array (TPA) programme paper series \citep{Johnston2020}.

\section{Observations and Analysis}
\label{sec:obs}

\begin{table*}
     \caption{MeerKAT PSR J0540$-$6919 data across three observing epochs. Here Nants, Nchan, BW and Npol refer to the number of MeerKAT antennas used during the observation; the number of frequency channels recorded; the associated MeerKAT bandwidth in MHz; and the number of recorded Stokes polarisation states respectively. Mean flux density values have estimated error bars of 15\%.}
    \label{tb:data}
    \begin{tabular}{lp{2.0cm}clllllllll}
    \hline
         Label&Date & Nants& Start\,time &Duration& Rotations& Nchan& BW & Npol & Mean flux density\\
         &&&(UTC)  & (sec) & &&MHz&& ($S_{\rm{Int}}$, mJy)\\
         \hline
         \\
         Epoch 1&26 March 2019 & 56 &16:36:59& 7200& 133188 &912&762&1&$0.13\pm 0.02$ \\
         Epoch 2&16 April 2019 & 56 &17:16:24& 7196& 135237 &768&642&4&$0.07 \pm 0.01$\\
         Epoch 3&27 August 2019& 61 &23:35:47& 7202& 135951&768&642&4&$0.10\pm 0.02$\\
         \\
         Totals&&&&21598&\multicolumn{3}{c}{404376 \,\, (5.7 hrs on source)}\\
\\
         \hline
    \end{tabular}
\end{table*}

\subsection{MeerKAT data}\label{sec:data}
Data for PSR J0540$-$6919 were recorded with the 64-element MeerKAT interferometer across three 2hr-observations during 2019. Hereafter, for brevity's sake, referred to as epoch 1 to 3. The observational parameters are provided in Table \ref{tb:data}. All data were recorded using the Meertime PTUSE back-end working in filterbank mode as described in \citet{bailes2020}. This mode records search-mode files with a time resolution of 38.3\,\si{\us} in the PSRFITS file format \citep{psrchive}. The MeerKAT band (856\,MHz -- 1712\,MHz) is split into 1024 coherently dedispersed channels. Epoch 1 and 3 were coherently dedispersed at a DM value of 146.5\,pc\,cm$^{-3}$. However, during epoch 2 the DM value was not set and therefore no coherent dedispersion was performed. The DM-smear in the central frequency channel at this DM value is equal to 480.567\,\si{\us}, less than the scattering exhibited by the pulsar.

Updates to the PTUSE back-end from March to August resulted in different data characteristics as follows. Epoch 1 data have a recorded bandwidth of 762\,MHz, while for epoch 2 and 3 data the bandwidth is 642\,MHz at the same centre frequency.  The epoch 2 and 3 data have full Stokes information, while epoch 1 data were recorded capturing total intensity only. In total we have single pulse data (recorded rotations) covering 5.7~hrs of observing. All these data predates automatic MeerKAT pipeline polarisation calibration as described in \citet{bailes2020}. 

\subsection{Data reduction and calibration}\label{sec:datared}

Single-pulse search-mode PSRFITS files were converted to pulsar archives by using an up to date X-ray ephemeris obtained from the Neutron star Interior Composition Explorer (NICER, Marshall, F. \textit{private comm.}) and \textsc{dspsr}\footnote{\href{http://dspsr.sourceforge.net/}{http://dspsr.sourceforge.net/}} software \citep{dspsr} to write out dispersion corrected archives for each pulsar rotation. The resulting data have 1024 phase bins across a pulse period, providing a phase bin resolution of 49.7\,\si{\us}. 

The observations were performed without flux calibrators. Therefore flux density values are estimated based on the modified radiometer equation \citep{pulsarhandbook} and the system noise statistics.

The \gls{sefd} of the array was calculated by scaling the SEFD of a MeerKAT antenna by the number of antennas used in each of the observations. Following commissioning work that reports the MeerKAT tied-array-beam to be consistently more than 95\% coherent, we assume a fully coherent beam \citep{Geyer2021}. The SEFD used is reported in online MeerKAT technical documentation\footnote{\href{https://science.ska.ac.za/meerkat}{https://science.ska.ac.za/meerkat}} and can be modelled by,
\begin{align}
    \rm{SEFD} &= 5.71\times 10^{-7}\,\nu^{3} - 1.90\times 10^{-3}\,\nu^{2}  + 1.90\,\nu  - 113,\label{eq:sefd}
\end{align}
with $\nu$ the observing frequency in MHz and the SEFD in Jy. 
The band averaged SEFD is $427 \pm 23$\,Jy, where the error represents deviations across antennas.  This leads to an array SEFD of 7.6~Jy for epochs 1 and 2 (56 antennas) and 7.0~Jy for epoch 3 (61 antennas).

Error bars on all computed mean flux densities in this paper are estimated to be 15\%, calculated from the above SEFD standard deviation and increased to accommodate systematic errors.

Polarisation calibration solutions were obtained offline as follows. During the array phase-up procedure the noise diodes on the antennas in the observing array were activated and the time and phase offsets between the data streams from the linearly polarised feeds computed. Additionally, a well known calibrator (PKS J1939--6342 or PKS J0408--6545) was observed to solve for individual antenna gain values and bandpass shapes. These computed solutions, stored in the observational metadata, were collated into a single polarisation calibration solution file which is then applied to the beamformed data using \texttt{pac} in \textsc{psrchive} \citep{psrchive}. More detailed information on the polarisation calibration procedure can be found in \citet{serylak2020arxiv}.

\subsection{Finding single pulses}\label{sec:searchpipeline}

We searched the single-rotation archive files for bright pulses to analyse as giant pulse candidates. To increase the efficiency of the searches the full-band data were integrated to 256 channels and 512 phase bins before boxcar searches were performed. 
This also improves the ease of \gls{rfi} identification and excision. RFI masks, unique to every 30~min of observing, removed between 16\% and 29\% of the frequency channels. The discovered band-limited nature of some of the giant pulses (see Sec.~\ref{sec:bl}) encouraged us to use \gls{rfi} masks that remove as few channels as possible during the search process.

After RFI removal the \gls{snr} values of each frequency averaged total intensity (Stokes I) single pulse file was computed using \texttt{psrstat} in \textsc{psrchive} \citep{psrchive}. As in \citet{Abbate2020}, a \gls{snr} cut-off of 7 was used to identify bright single pulses.  The average \gls{snr}, computed from the \gls{snr} of the integrated profile divided by the square root of the number of rotations observed, is found to be less than 0.2. As such, a single pulse with \gls{snr} = 7 represents a clear deviation, and is unlikely to be confused with emission following a log-normal distribution.

Finding bright single pulses was an iterative procedure, whereby RFI masks were updated if the pipeline produced many (>100) false positive candidates in quick succession. The resulting bright single pulses were inspected by eye for validity. 

\section{Results}\label{sec:results}

\subsection{Integrated profile for PSR J0540$-$6919}\label{sec:aveprof}

\begin{figure}
\centering
\includegraphics[width=0.9\columnwidth]{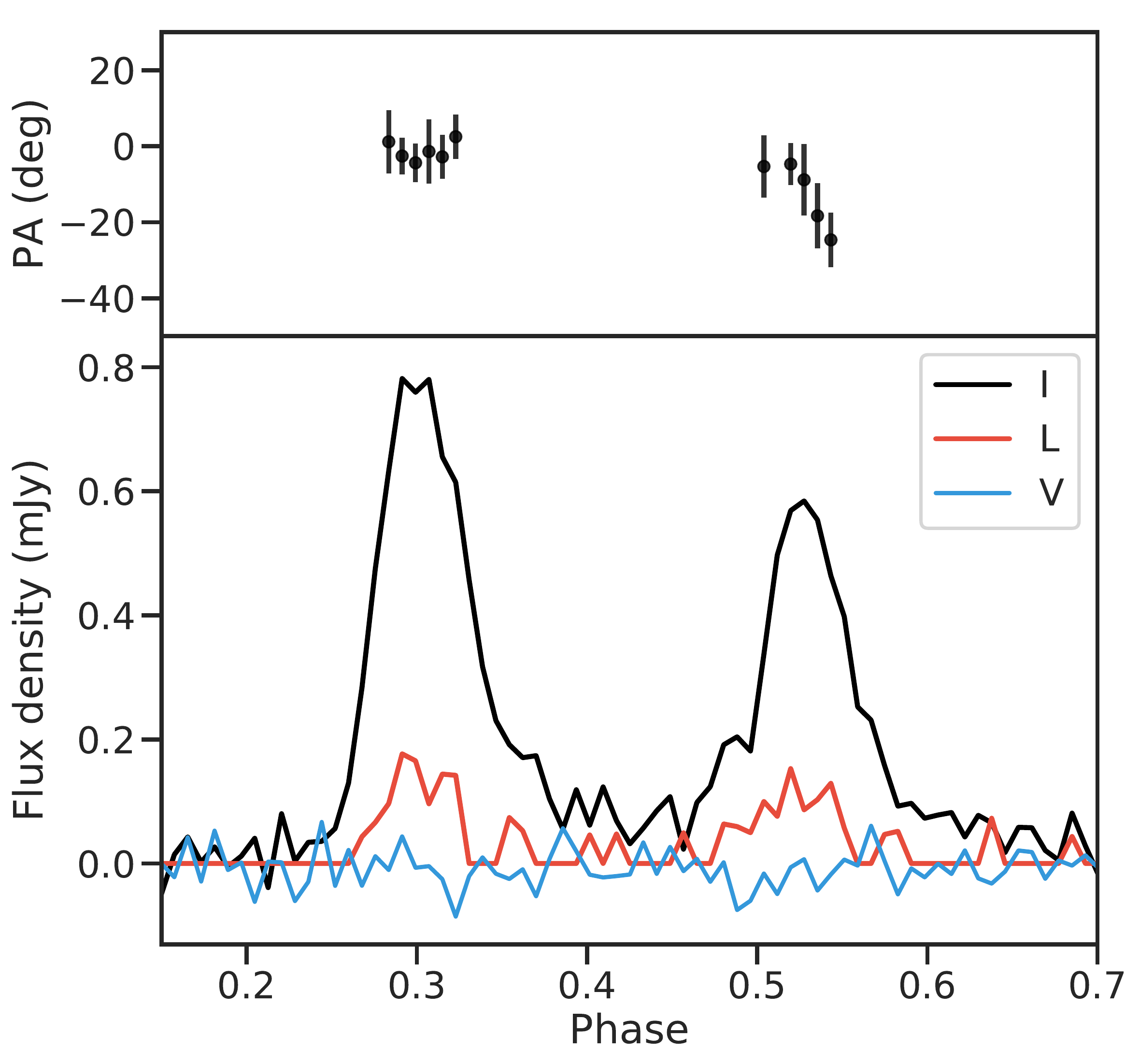}
\caption{The integrated pulse profile for epoch 3 averaged down to 128 phase bins to present its mean polarisation properties. A similar integrated profile was obtained for the other epochs, however for epoch 1 we do not have polarisation information. The estimated mean flux densities at each epoch are presented in Table \ref{tb:data}. This profile has been RM corrected by our best RM estimate of -245.8 as computed in Sec.~\ref{sec:rm}. Position angles are plotted for bins for which the linear polarisation has a \gls{snr} > 3.0, with respect to the Stokes I off-pulse RMS.} \label{fig:ave}
\end{figure}

Integrated profiles for each epoch were obtained and calibration applied as in Sec.~\ref{sec:datared}. Fig.~\ref{fig:ave} shows the integrated pulse profile for epoch 3 $-$ the epoch with the brightest integrated profile with full polarisation information. Two distinct windows of emission are clearly visible and separated by approximately a quarter of a pulse rotation: the first is centered at 0.3 (leading pulse) and the next at 0.53 (trailing pulse). The peak flux densities of the leading pulse is consistently higher than the trailing pulse for all epochs. The mean flux densities measured on the three epochs range from 0.07\,mJy to 0.13\,mJy, with the peak flux densities ranging from  0.55\,mJy to 0.87\,mJy.

The biggest change was observed between epoch 1 and 2 where the mean flux density almost doubles within three weeks,  and is likely due to refractive scintillation (e.g. \citealt{rnb86}). The changes in observed flux densities also scale with the number of bright single pulses found per epoch. From our three mean flux density measurements we estimate a modulation index of $\sim$0.3, computed from the standard deviation of the flux values centered on the leading (or trailing) pulse divided by the associated mean value \citet{Jenet2003}. A dedicated campaign of regular observations would be required to analyse these changes in detail and more accurately compute the modulation index. We measure a steep spectral index of $-3.2 \pm 0.3$ for the integrated profile, averaged across eight frequency bands.

The DM values at which the average profiles are dedispersed range from 146.9 to 147.1 pc cm$^{-3}$, differing from the original coherent dedispersion parameter of 146.85 pc cm$^{-3}$. These DM values were obtained by maximising single pulse \gls{snr} as these have higher signal-to-noise ratios than the mean pulse profile. Similarly, the best RM estimate was computed from the brightest GPs, as described in Sec.~\ref{sec:rm}.

Fig.~\ref{fig:ave} also displays the Stokes V component, the linear polarisation (L), and the position angles for bins where $L$ has a \gls{snr} > 3.0  with respect to the Stokes I off-pulse \gls{rms} value. The integrated profile has a linearly polarised fraction of only 28\%, and a distinct position angle swing across the trailing emitting phase window. 

\subsection{The Giant Pulse population}\label{sec:stat}
\begin{table}
    \caption{The number of GPs found for each MeerKAT observing epoch. The number of GPs discovered in the leading (L) or trailing (T) phase emission window are indicated in brackets in column 2. 
    Columns 3 to 5 show GPs classified according to the categories of Sec.~\ref{sec:indv}.  
    The averaged GP mean flux density, $S_{\langle\rm{GP}\rangle}$, per epoch is presented in the last column. }
    \label{tb:giants}
    \begin{tabular}[width=\columnwidth]{llllll}
    \hline
         Label&Nr of GPs&GPs \gls{snr}&Band-lim.&Double &$S_{\langle\rm{GP}\rangle}$\\
         &(L/T)&>100&GPs&GPs & (mJy) \\
         \hline
         \\
         Epoch 1&  441 (229/212)&1&16&3&3.7\\
         Epoch 2&  130 (65/65)   &0&2&0&3.0 \\
         Epoch 3&  294 (154/140)&2&6&2&3.2\\
         \\
         Total& 865 & 3 & 24 & 5\\
\\
         \hline
    \end{tabular}
\end{table}

We discovered a total of 865 bright single pulses in 404376 rotations. The highest number of bright single pulses were found in epoch 1 (441), followed by epoch 3 (294) and epoch 2 (130), as presented in Table \ref{tb:giants}. The high occurrence rate of epoch 1 corresponds to one bright single pulse (\gls{snr} > 7) on average every 16\,s or correspondingly every 315 rotations. This is more than a hundred-fold increase in detection rate compared to approximately two GPs per hour published in \citet{jrmz04}. 

\subsubsection{Giant pulse flux density distribution}\label{sec:fluxdistr}

\begin{figure*}
    \centering
    \includegraphics[width=1.7\columnwidth]{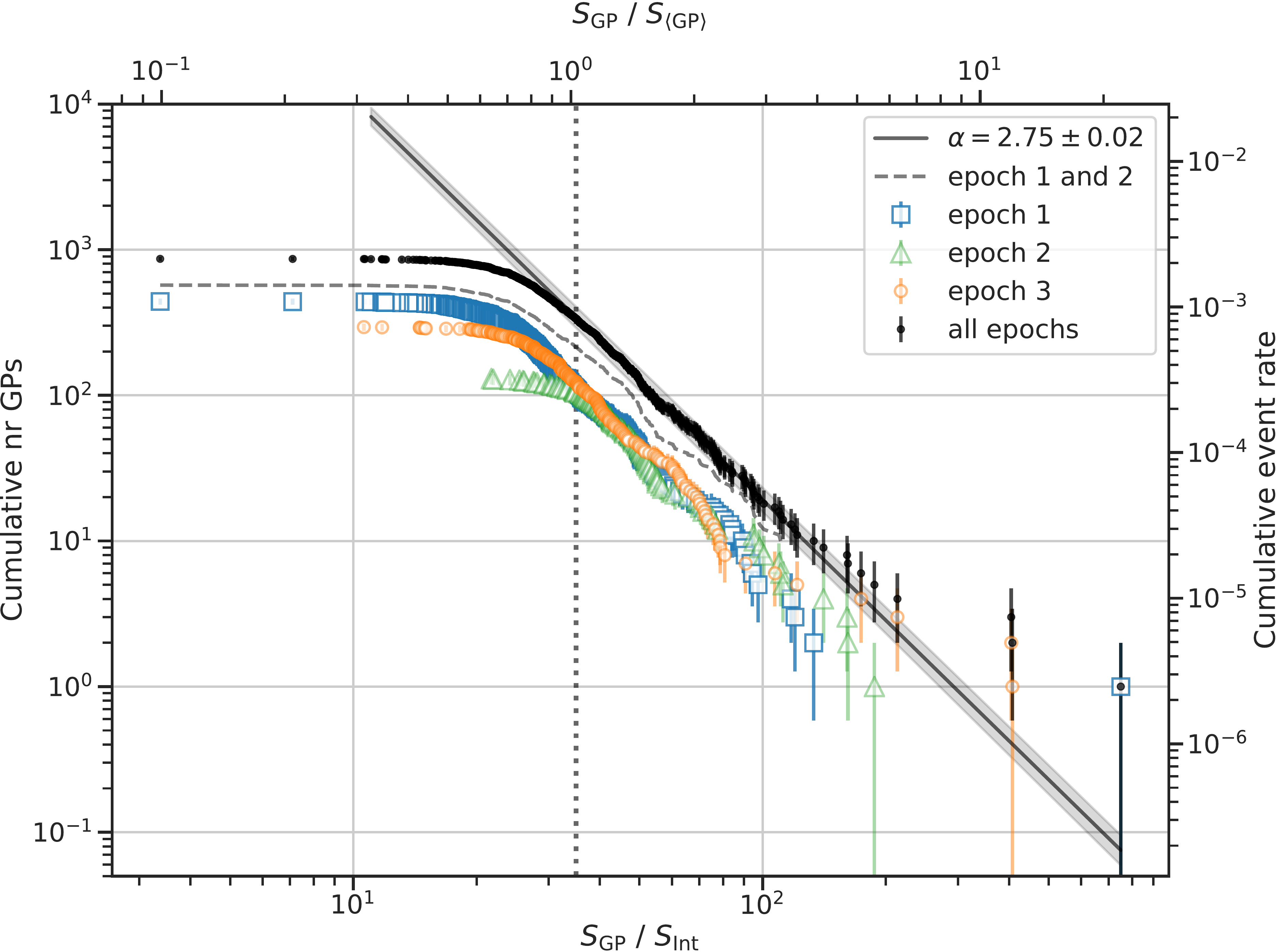}
    \caption{The flux density distributions of epochs independently (in colour) and all three epochs combined (black). The GP mean flux densities ($S_{\rm{GP}}$) are normalized by the flux density ($S_{\rm{Int}}$) to account for epoch-to-epoch variation, and plotted against the cumulative number of GPs. Error bars are equal to the square root of the cumulative sample size of the GPs. A best fit power law to the combined data set, with $S_{\rm{GP}} > 35\times S_{\rm{Int}}$ (dotted line), has a power law index of $-2.75 \pm 0.02$. The $1\sigma$ error region to this best-fit is shown in grey (shaded). Secondary axes show the cumulative event rate for a GP of a given flux density (right), and the flux densities expressed in units of the averaged GP flux density ($S_{\langle\rm{GP}\rangle}$, top axis). These axes relate only to the combined epoch's data points. Local deviations from the power law slope (`kinks') are seen in the data of all three epochs. These kinks most prominently at $\sim 50\times S_{\rm{Int}}$ line up particularly well for the distributions of epoch 1 and 2 that were observed only three weeks apart. Their combined distribution emphasising these kinks is shown as a dashed curve (for cumulative number of GPs > 10).}
    \label{fig:dist}
\end{figure*}

The single pulse mean flux density distribution is shown in Fig.~\ref{fig:dist} and provides evidence that the discovered bright single pulses do follow a power law distribution, consistent with 
other pulsars that emit GPs.

The flux calibrated distribution for each epoch was obtained by calibrating the single pulse dynamic spectra using eq. \eqref{eq:sefd} and a constant off-pulse \gls{rms} per epoch, and thereafter averaging across the full MeerKAT band to produce flux calibrated profiles. 

To compensate for the observed differences in epoch mean flux densities ($S_{\rm{Int}}$) we normalized the distributions of GP mean flux densities ($S_{\rm{GP}}$) of each epoch independently. The best-suited normalizing values which aligned the three distributions, were found by sampling values of $S_{\rm{Int}}$ within the 15\% error bars presented in Table~\ref{tb:data} (in steps of 0.001\,mJy). We found that choosing $S_{\rm{Int}}$ equal to 0.132, 0.071 and 0.100\,mJy, respectively, aligned the distributions well.  In Fig.~\ref{fig:dist} these values were used as the normalization factors to produce the cumulative GP distributions for each epoch (in colour). The combined flux density distribution is also shown (in black) and normalized by an averaged  $S_{\rm{Int}}$ of 0.1\,mJy. For the combined epoch data we also present the cumulative event rate for a GP of a given flux density on the right axis (in terms of the total number of rotations observed).

As in \citet{jrmz04} the distributions appears to follow a power law dependence, expressed in the form
\begin{equation}
\rm{P}(S>S_0) = K\,\,S_0^{-\alpha}, \label{eq:prob}
\end{equation}

\noindent with $K$ and $\alpha$ the amplitude and power law index respectively, and $S_0$ the normalized GP flux density ($S_{\rm{GP}}/S_{\rm{Int}}$), such that $\rm{P}(S>S_0)$  is the rate of occurrence of GPs above $S_0$.

A flattening of data points towards the lower end of the flux distribution for each epoch indicates that our GP data set is likely incomplete for low flux density values. This suspicion is confirmed by observing that the GPs exhibit a large distribution in widths, such that a cut in peak \gls{snr} (as used to search for bright pulses,  Sec.~\ref{sec:searchpipeline}) does not translate to a uniform cut in mean flux density. 
Examining the $S_{\rm{GP}}$ distribution of pulses with \gls{snr} < 7.3 we find that the set is complete for mean flux density values > 3.0\,mJy. Therefore all power law fits are applied only to GPs for which $S_{\rm{GP}} > 3.0$\,mJy or conservatively above 35$\times$\,$S_{\rm{Int}}$ (dotted line in Fig.~\ref{fig:dist}). 

The best power law fits to the data are obtained using a Bayesian chi-square minimization algorithm based on the \texttt{EMCEE}\footnote{\href{https://github.com/dfm/emcee}{https://github.com/dfm/emcee}} python package \citep{Foreman-Mackey2013}. Single power law fits for all epochs have similar power law indices ($\alpha$), the estimates of which are $2.79 \pm 0.09$ (epoch 1), $2.79 \pm 0.08$ (epoch 2) and $2.79 \pm 0.07$ (epoch 3). The power law index obtained by fitting the combined distribution is $2.75 \pm 0.02$. 

Although the general trend is well described by a single power law, we observe local deviations from the power law slope, or `kinks' in the data, for all three epochs. For epochs 1 and 2 these kinks line up particularly well (after normalizing by $S_{\rm{Int}}$) at about 50 to 60$\times$\,$S_{\rm{Int}}$.

The power law best-fit is seen to deviate from the measured data points towards the higher end of the flux distribution.  With limited data at the tail end of the distribution we fail to constrain a power law index for $S_{\rm{GP}} > 15$\,mJy (or 150$\times S_{\rm{Int}}$). At these values the error bars equal to the square root of the number of cumulative events also become proportionally large. We caution that with only a few data points at large flux values the supposed deviation is not conclusive.

By the description of our power law fit the cumulative event rate for a pulse at least as bright as the brightest pulse across all epochs (Fig.~\ref{fig:gallery}a) is 1.87$\times$10$^{-7}$ (Fig.~\ref{fig:dist}). Following the Poisson waiting time description in Appendix \ref{sec:app}, eq. \eqref{eq:app3}, more than 12 million pulse rotations are required to observe a pulse this bright with a 90\% probability. This is reduced to > 7 million pulses using the upper limit (95\% confidence, shaded region of Fig.~\ref{fig:dist}) of the best-fit power law to extract the cumulative event rate of the brightest GP. Alternatively, using the cumulative event rate of the data point value itself (2.47$\times$10$^{-6}$), the waiting time for such a bright pulse with 90\% confidence is > 930\,000 rotations. Depending on whether the power law description continues unbroken to high flux density values, it would therefore require between 2 and 20 to 30 equivalent observing campaigns (of 5.7 hrs) to observe a GP at least as bright as our brightest GP with a 90\% probability.

The averaged $S_{\rm{GP}}$ across all detected 865 GPs is $S_{\langle\rm{GP}\rangle} = 3.4$\,mJy. 
The top axis of Fig.~\ref{fig:dist} expresses $S_{\rm{GP}}$ in units of this mean. Compared to a mean flux density of $S_{\rm{Int}} = 0.1$\,mJy, averaged across the three epochs, the 865 GPs make up only 7\% of the detected flux ($\sim0.1\,\rm{mJy}\times404376$). The $S_{\langle\rm{GP}\rangle}$ estimate and its fractional contribution to the overall flux would change to 3.9\,mJy and 8\% when using only the GPs that form part of the complete set ($S_{\rm{GP}} > 3.0$\,mJy).

\begin{figure}
    \centering
    \includegraphics[width=\columnwidth]{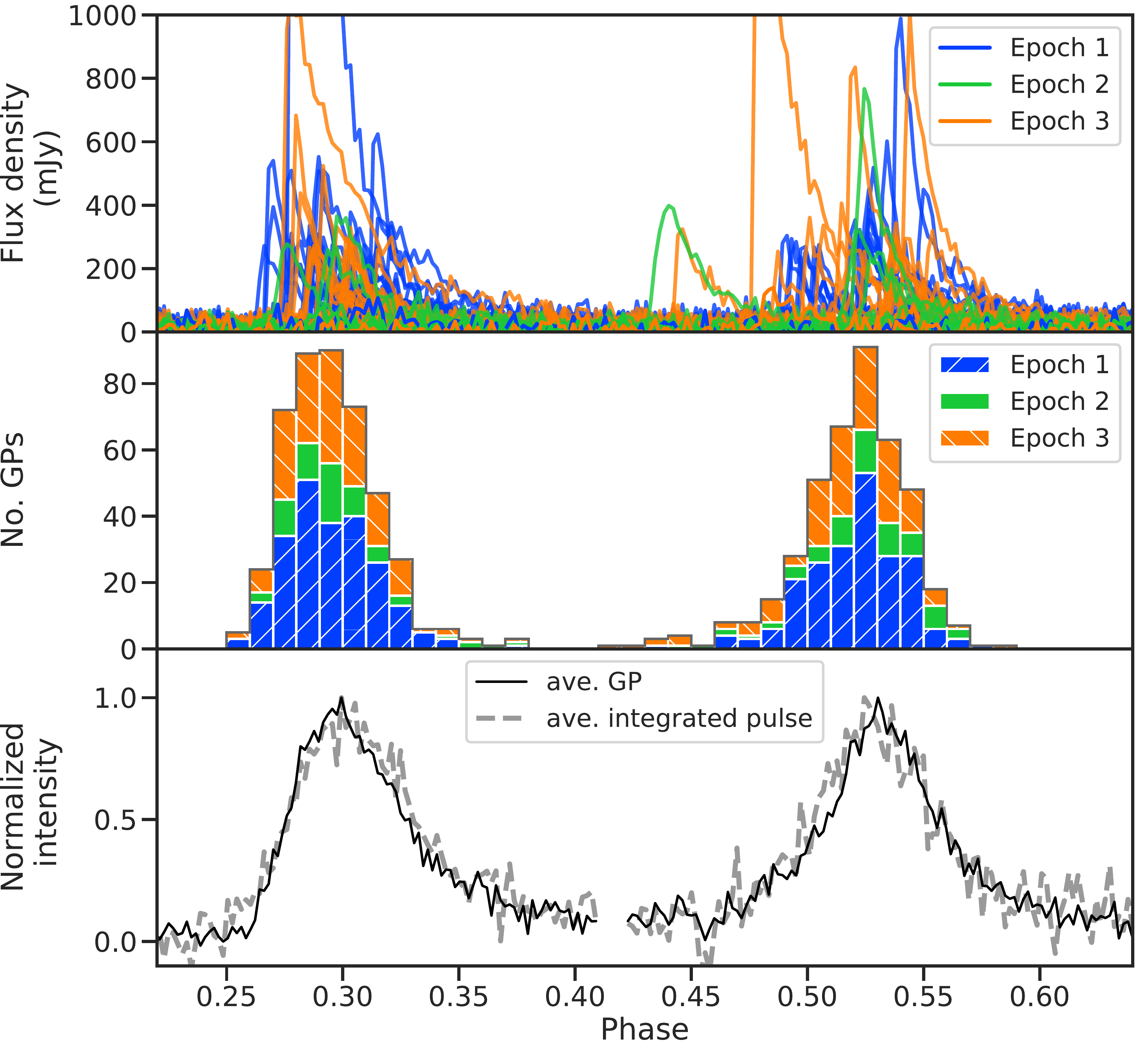}
    \caption{\textit{Top}: Profiles of all the GPs with peak flux densities > 250\,mJy. The flux density axis is artificially cut at 1~Jy so that the very bright GPs do not prohibit examination of the weaker pulse shapes. \textit{Middle:} A stacked histogram of the phases for all GPs. \textit{Bottom:} Comparing the average GP pulse shape with the average integrated profile. The plotted peaks have been normalized to unity for each emission window (leading or trailing) independently. The break in the profile after 0.4 phase serves to indicate this change in normalization factor.}
    \label{fig:allgps}
\end{figure}

\subsubsection{Giant pulse phase distribution}

The detected GP population exhibits a bi-modal phase distribution, as seen in the top panel of Fig.~\ref{fig:allgps} for each epoch. There is significant variation in the phases at which the GPs occur, the stacked histogram of which is shown in the middle panel. 

From the 865 giants, 448 were are detected in the leading pulse phase window and 417 in the trailing phase window (see Table~\ref{tb:giants} for epoch breakdowns), such that there is at most a 10\% difference in the number of GPs emitted from the two phase windows. The power law indices for treating the leading or trailing GP flux distributions per epoch independently lie within 1$\sigma$ of one another. This shows that leading and trailing GPs have very similar flux density distributions.

We find that for each epoch the mean GP profile shape is similar to the average integrated profile. In the bottom panel of Fig.~\ref{fig:allgps} the similarities between the averaged GP shape (across all detected GPs), and the averaged integrated profile (across all epochs) can be seen. Note that in order to make these comparisons we have normalized the intensity of the peaks of the averaged GP and integrated profile to unity in each emission phase window (leading and trailing) independently. 

\subsubsection{Giant pulse arrival times}\label{sec:toa}

\begin{figure}
    \centering
    \includegraphics[width=0.9\columnwidth]{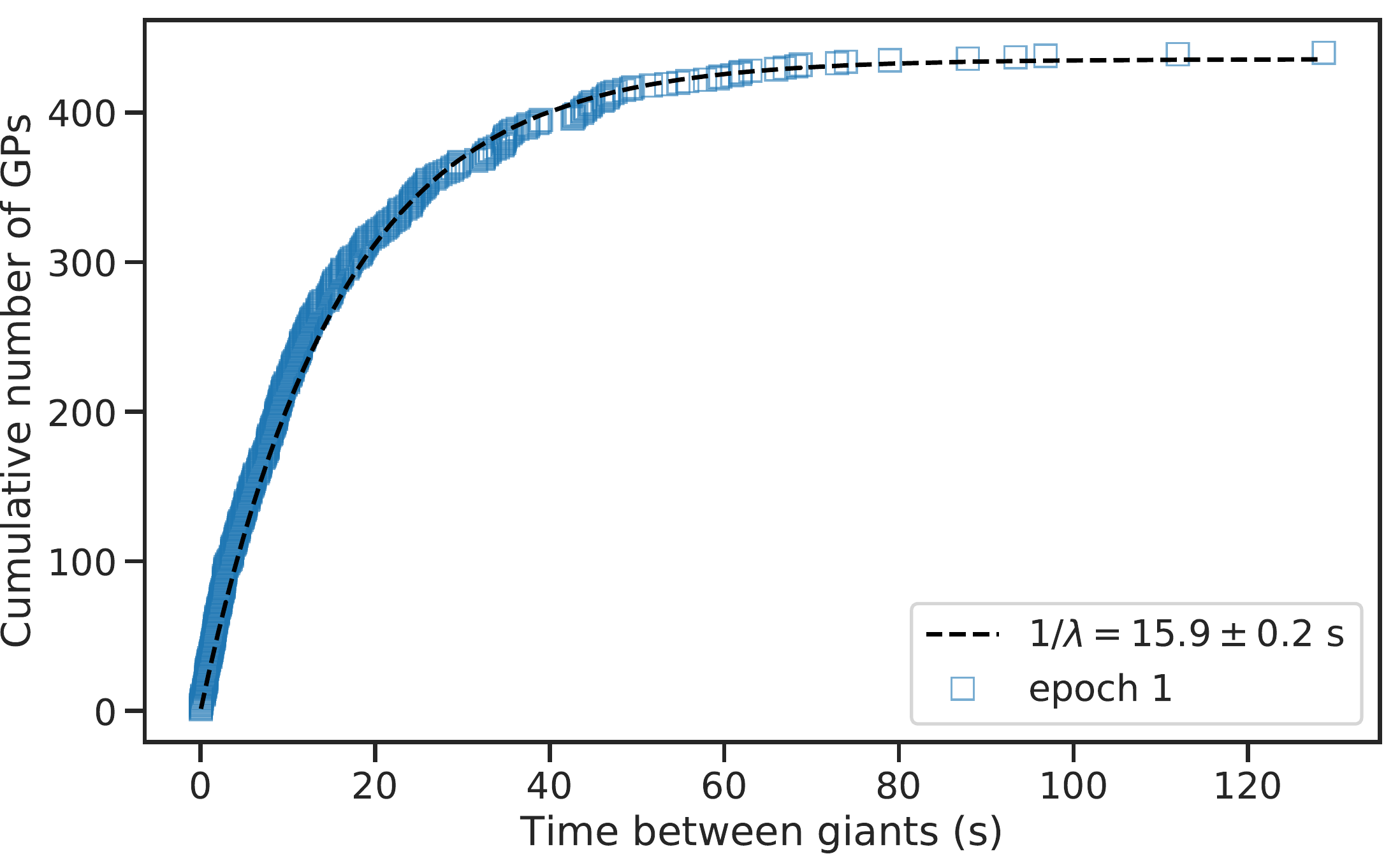}
    \caption{The cumulative distribution of arrival times between detected GPs in epoch 1.  The fitted exponential function ($1-e^{-\lambda t}$, dashed line) has a mean waiting time of $15.9 \pm 0.2$\,s.}
    \label{fig:times}
\end{figure}

To investigate whether the detected GP sample represents emission independent of one another, we compare the GP arrival times to a Poisson distribution. Data are effectively Poissonian if we can show that the cumulative distribution of measured delays, or waiting times, between GPs follows an exponential function proportional to  $1-e^{-\lambda t}$, with $\lambda$ the GP event rate (1/waiting time) as detailed in Appendix \ref{sec:app}. 

We find that the waiting times between GPs in epoch 1 are well fit by the above exponential form with $1/\lambda = 15.9 \pm 0.2$\,s, as presented in Fig.~\ref{fig:times}. Alternatively $\lambda$ can be computed from the mean ($1/\lambda$) and median ($\ln2/\lambda$) values of an exponential function. The mean and median values of the waiting time distribution provide $1/\lambda=14.5$\,s and $16.3$\,s respectively, encompassing our best-fit value. 

We test the null hypothesis that the observed distribution of wait times is drawn from an exponential distribution using the Kolmogorov-Smirnov (KS) test. For the full distribution we find an associated p-value statistic >~0.2 and therefore cannot reject this null hypothesis. 

We also test subsets of the waiting time distribution (with decreasing maximum waiting times and number of data points) against a truncated exponential function (exponential function multiplied by a unit-step function), and find that the associated KS-test p-value remains > 0.02 for all small time distributions; and > 0.1 for waiting time distributions with a maximum waiting time above 50\,s. Furthermore, compared to the p-values obtained from testing for a uniform distribution (an alternate null hypothesis), we find that a truncated  exponential distribution is preferred over a uniform distribution for all distributions of smaller maximum waiting times. We therefore conclude that the overall distribution is well described by a Poisson point process, such that GP events on the whole are independent of one another. 

On even smaller time-scales we do find a rapid sequence of four GPs, as discussed in the next section (Sec.~\ref{sec:bl}, Fig.~\ref{fig:seq}), that could suggest GP clustering at < 1\,s time-scales.  However, below 3\,s the waiting time distribution decreases to a sample size $< 100$ rendering KS-test results increasingly uncertain for these even shorter time-scales.

\subsection{Individual giant pulse characteristics}\label{sec:indv}

\begin{figure*}
	\centering
	\begin{subfigure}[t]{\columnwidth}
		\centering
		\includegraphics[width=0.95\columnwidth]{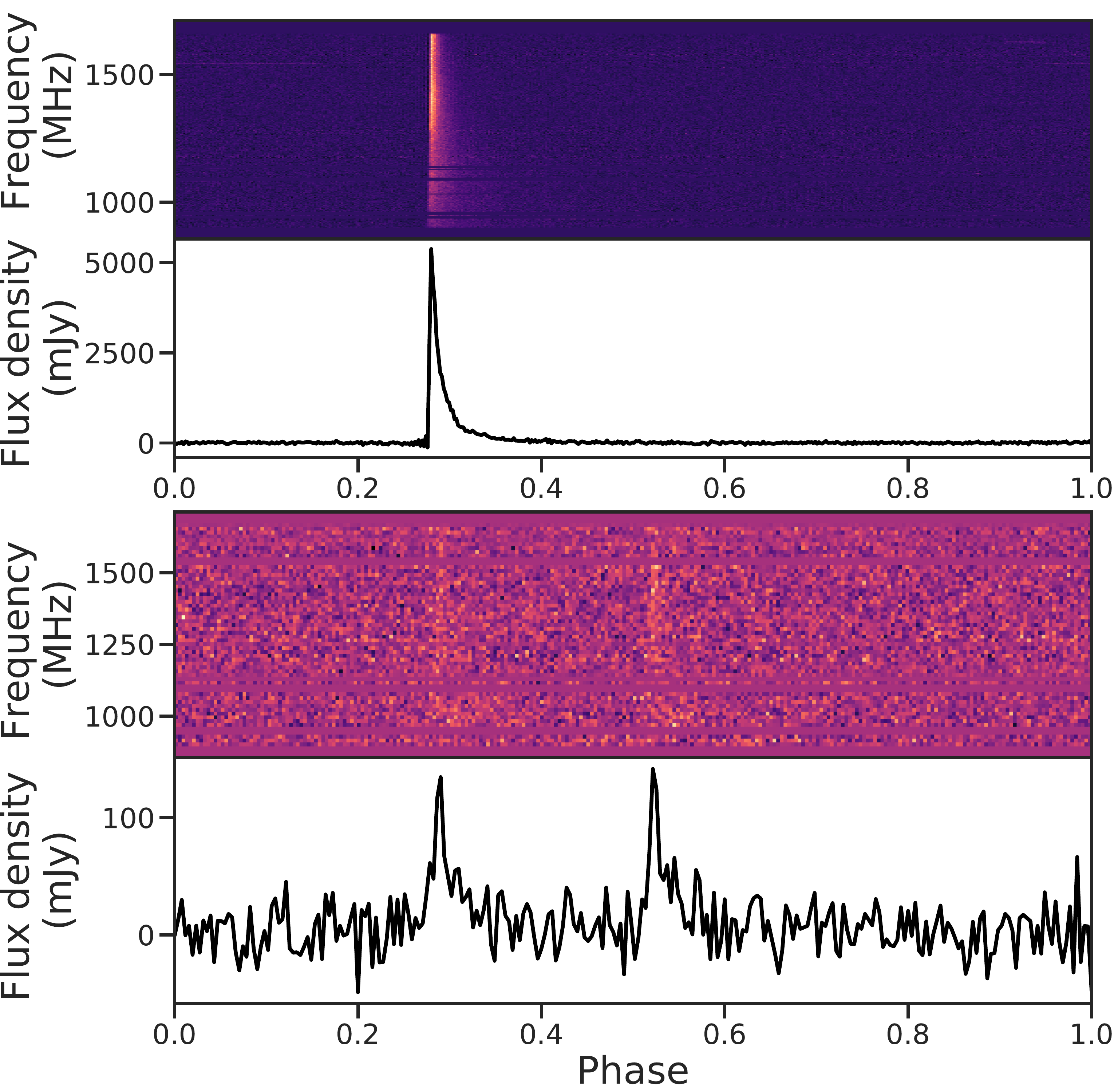}
		\caption{\quad\quad (a) top ; \quad (b) bottom} \label{fig:1ab}		
	\end{subfigure}
	\quad
	\vspace*{2em}
	\begin{subfigure}[t]{\columnwidth}
		\centering
		\includegraphics[width=0.95 \columnwidth]{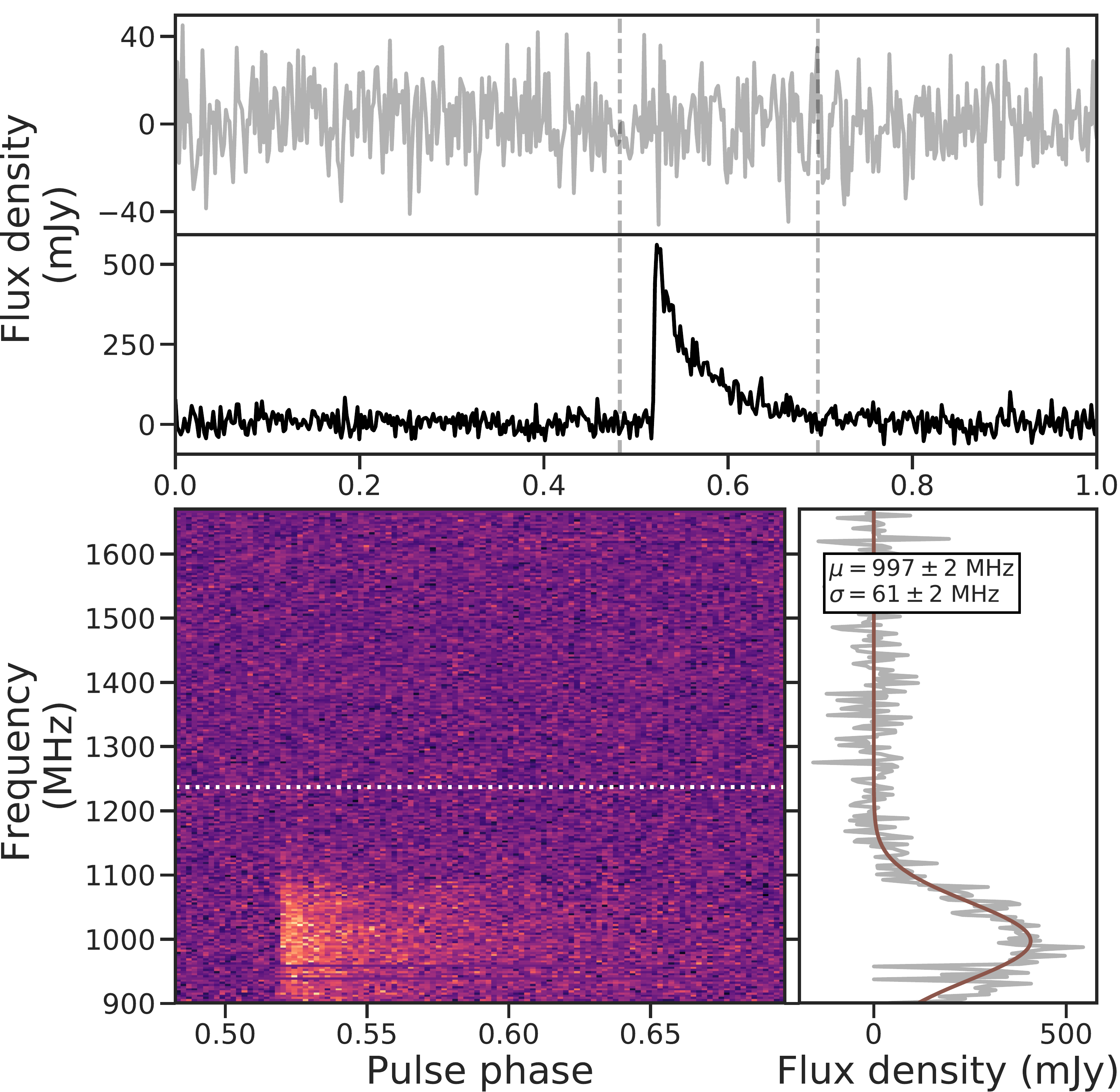}
		\caption{(c)}\label{fig:1c}
	\end{subfigure}

	\begin{subfigure}[t]{\columnwidth}
		\centering
		\includegraphics[width=0.95\columnwidth]{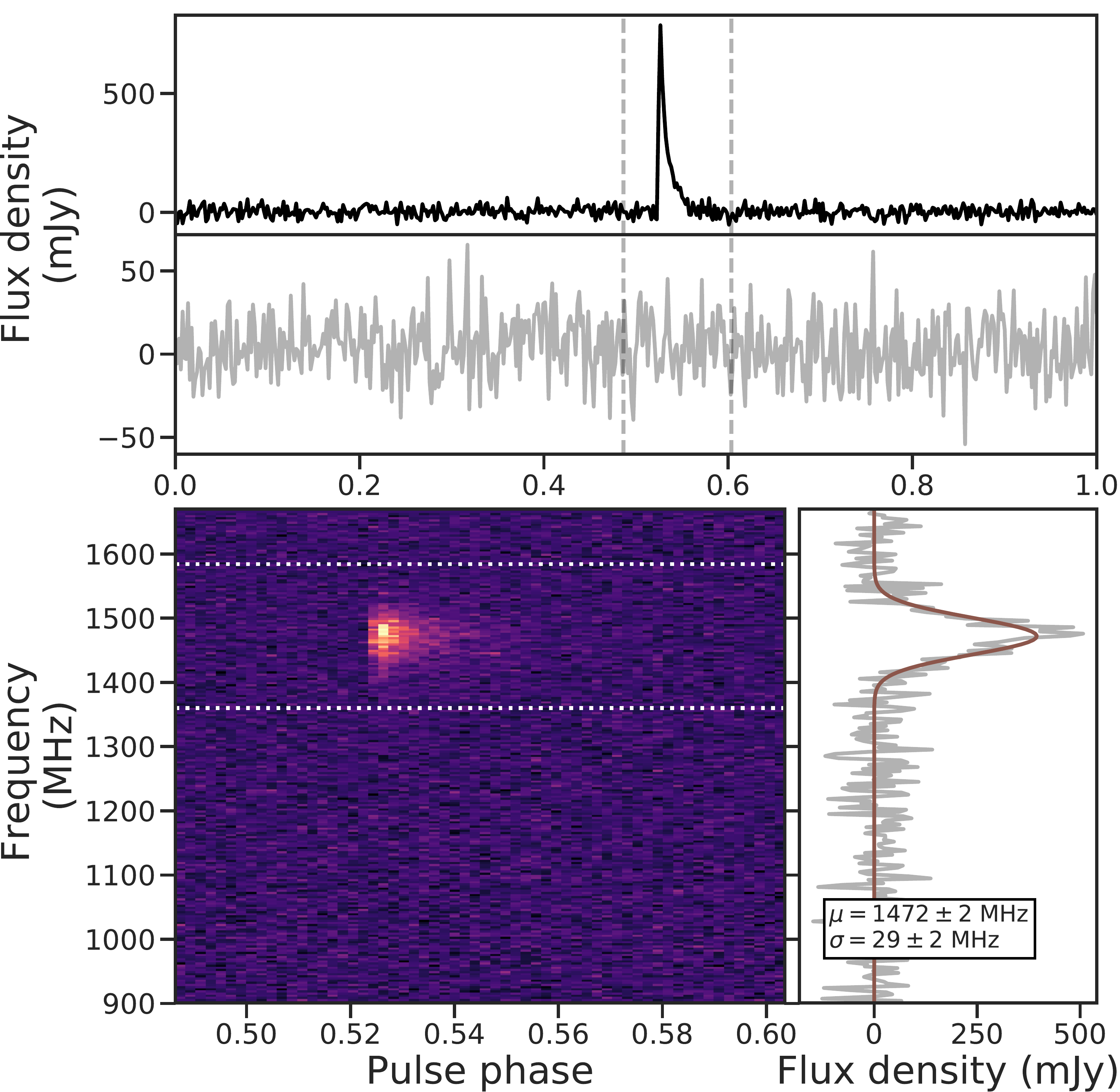}
		\caption{(d)}\label{fig:1d}		
	\end{subfigure}
	\quad
	\vspace*{2em}
	\begin{subfigure}[t]{\columnwidth}
		\centering
		\includegraphics[width=0.95\columnwidth]{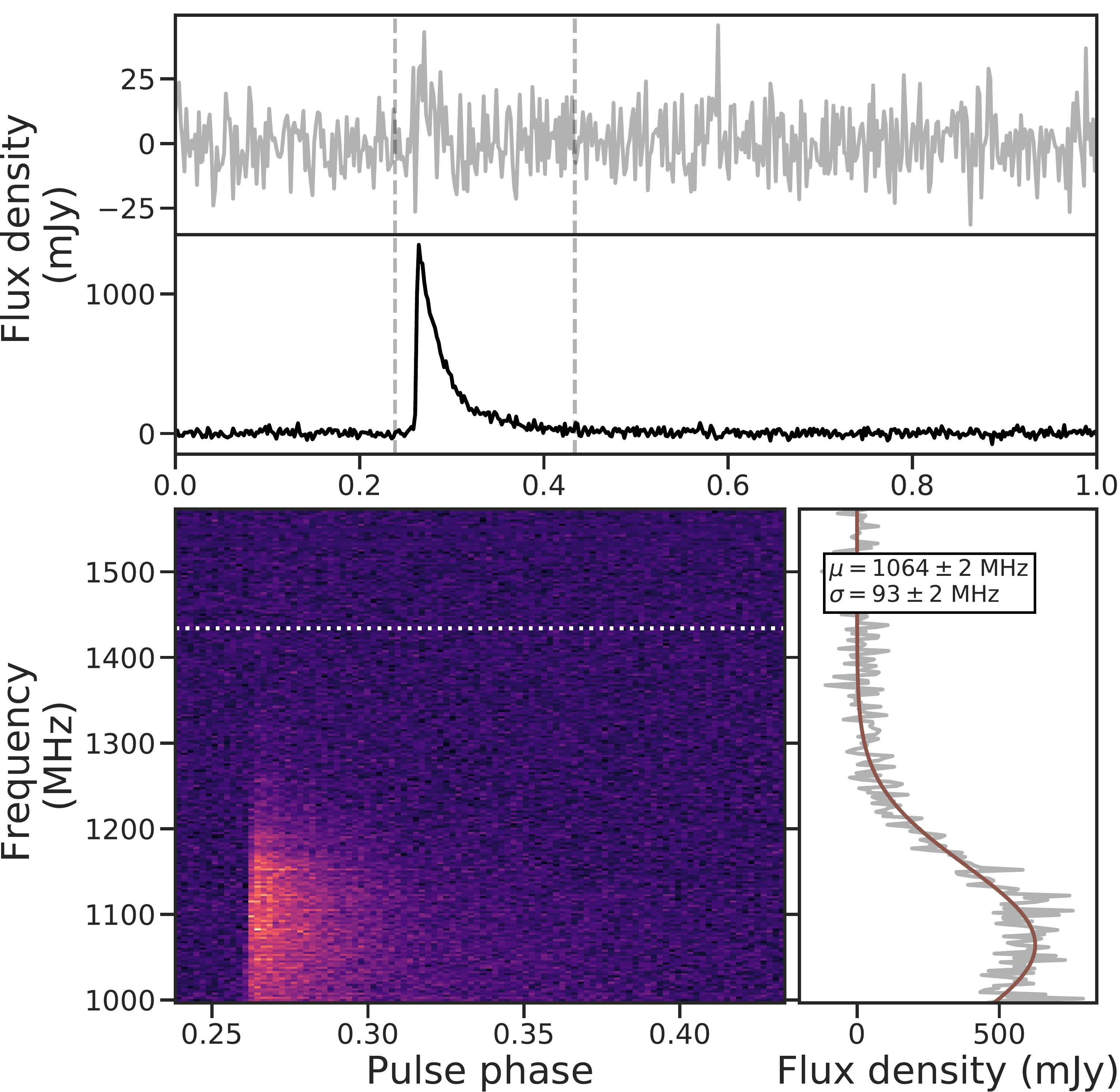}
		\caption{(e)}\label{fig:1e}
	\end{subfigure}	
    \caption{A selection of the detected GPs. Unless otherwise stated plotted resolutions are 256 frequency channels and 512 phase bins. \textit{(a)} At MJD 58568.7681 (UT 18:25:58.00) in epoch 1 we find a single pulse from the leading pulse window with a \gls{snr} > 400 (peak flux density of 6.2\,Jy) and an estimated mean flux density over one pulse period of 98.9\,mJy. \textit{(b})  The brightest ``double giant'' pulse found. For plotting purposes the double GP data are averaged to 64 frequency channels and 256 phase bins. \textit{(c) -- (e)} A selection of band-limited pulses and their corresponding frequency responses fit with a Gaussian model. The on-pulse frequency response is presented as a flux density in\,mJy that has been averaged over the number of on-pulse phase bins. On-pulse regions are designated by dashed lines in the top panels. The narrowest in frequency band-limited pulse is shown in \textit{(d)} and has a Gaussian $1\sigma$ width of 29 $\pm$ 2\,MHz. The dynamic spectra are split into frequency sections (dotted lines), and the averaged pulse profiles associated with these sections shown in the panels above it. In all but one case we are unable to detect emission beyond 4$\sigma$ from the peak frequency of bright flux knot. These $4\sigma$ separations are shown using dotted lines. The GP in panel $(e)$ is the exception, where the pulse profile is detected in the top part of the band as well, beyond the 4$\sigma$ line. This is the highest \gls{snr} band-limited pulse found, with a peak flux density of 1.4\,Jy.}\label{fig:gallery}
\end{figure*}

Here we describe the detected GPs in terms of:
the \textit{brightest giants} -- the very high end of the flux distribution; \textit{double giants} -- rare chance events of emission and \textit{band-limited giants} for which emission is detected in only a fraction of the observing band. Table \ref{tb:giants} provides an overview of the GPs found per epoch, and Fig.~\ref{fig:gallery} shows a collection of single pulse profiles.

\captionsetup[subfigure]{labelformat=empty}

\subsubsection{The Brightest giants}\label{sec:sg}

We briefly describe the three GPs with the highest \gls{snr} (>100) values as they are the best candidates to investigate other GP characteristics such as scatter broadening and polarisation levels. The largest \gls{snr} value measured across all epochs is for a GP towards the end of epoch 1 (MJD 58568.7681) with a \gls{snr}$> 400$, as shown in Fig.~\ref{fig:gallery}a. The peak flux density for this pulse is 6.2\,Jy, and the mean flux density of 98.9\,mJy (error bar of 15\%) is almost 800$\times$ the mean flux density for this epoch (0.13\,mJy). Using our smallest bin resolution of 0.05\,ms as the pulse width this evaluates to a fluence of 4.1\,Jy\,ms. At the distance to the Crab pulsar (taken here as 2.0\,kpc as in \citealt{Kaplan2008}) such a pulse would have a peak flux density 625$\times$ higher, i.e. 4\,kJy; and if we account for scattering (again assuming a conservative intrinsic pulse width of 0.05\,ms) a scatter-corrected peak flux density $\geq 10^{5}$\,Jy. This lower limit value is an order of magnitude smaller, compared to the brightest Crab GPs in \citet{he07} ($\sim$2\,MJy at 9.2\,GHz) and \citet{Bera2019} ($\sim$4\,MJy at 1.3\,GHz), but higher than the anomalously bright Crab GP of $\sim$155\,kJy at 430\,MHz in \citet{Cordes2004}, using a spectral index of -1.4 for the Crab main pulse \citep{Karuppusamy2010}.

In epoch 3 (at MJDs 58723.0411 and 58722.9895) we observe another two bright GPs each with \gls{snr}$ > 120$, and flux density estimates of 39.5\,mJy and 38.6\,mJy. In comparison, the largest mean flux density reported in \cite{jrmz04} across 72 hours of observing is 64\,mJy - which lies in-between our three brightest GPs. Since epoch~3 has full polarisation information, we investigate the polarisation properties and the rotation measure towards the \gls{lmc} using these bright GPs in Sec.~\ref{sec:rm}. Of particular interest is one of these bright GPs that shows band-limited emission (Fig.~\ref{fig:gallery}e) as described in the next section.

\subsubsection{Band-limited giants}\label{sec:bl}

\begin{figure}
    \centering
    \includegraphics[width=0.9\columnwidth]{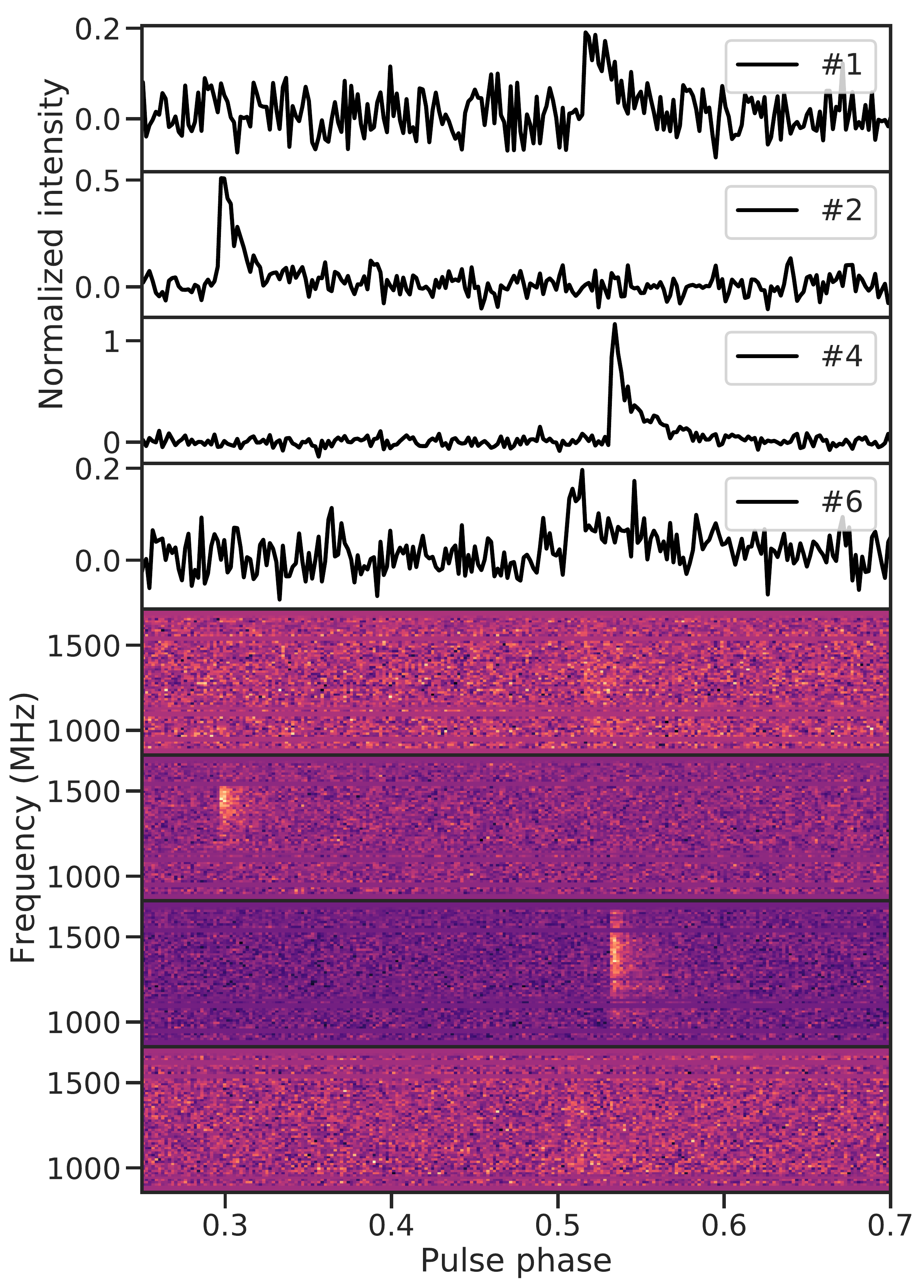}
    \caption{A rare sequence of GPs: in just six rotations four GPs are found, numbered \#1, \#2, \#4 and \#6. No GPs are detected for rotations \#3 and \#5. For plotting purposes the frequency axes have been integrated to 64 channels, and the phase bins to 512. The pulse profile intensities are plotted relative to the brightest \#4 pulse which is normalized to an intensity value of 1. In just two pulse rotations the GPs change from highly band-limited in pulse \#2 to broad-band in pulse \#4.}
    \label{fig:seq}
\end{figure}

Band-limited pulses of varying bandwidths were discovered during each epoch. These are GPs for which the radio emission appears to not cover the full MeerKAT band. Across all epochs a total of 24 band-limited giants were found. The band-limited giants were searched for by inspecting the integrated on-pulse spectrum (frequency response) of all the discovered giants. GPs for which a simple Gaussian fit across the integrated on-pulse spectrum had a standard deviation width, $\sigma$ < 200\,MHz were selected for further inspection. Most band-limited pulses found in this way were simultaneously easily identified by eye. To improve the Gaussian fits across the on-pulse spectra minimal RFI excision was used on identified candidates. 

These band-limited pulses appear as `flux knots': single bright regions of emission in the associated frequency-time plots, that cover only a fraction of the band and that are visibly symmetrical in their frequency response. Of the 24 flux knots identified, 18 had $\sigma$ < 100\,MHz. The band-limited nature of these flux knots were checked by removing its emission signature (beyond 4$\sigma$ of the peak frequency) and confirming that the remaining data did not contain a buried broad-band signal. This was found to be true for all but one GP (the brightest, Fig.~\ref{fig:gallery}e) with $\sigma$ < 100\,MHz. 

The narrowest flux knot has a Gaussian 1$\sigma$ frequency width of 29 $\pm$ 2\,MHz.  The GP peak flux density associated with 140\,MHz covered by the flux knot (or $5\sigma$ across) is $\sim$940\,mJy, a two-fold increase in peak value compared to integrating the full MeerKAT band. The broadest flux knot spans approximately half the observing band, with $\sigma = 184 \pm 16$\,MHz. The mean $\sigma$ value across the 24 band-limited GPs is 84\,MHz. Across the distinctly narrow knots ($\sigma$ < 100\,MHz) the mean value drops to 62\,MHz (or a full width half maximum value of 146\,MHz). 

Band-limited pulses occurred in different parts of the band, i.e both in the upper and lower sections of the band, but no flux knot was found to have its peak frequency between 1180 and 1325\,MHz. This frequency-window does however overlap with sections typically badly affected by RFI (see Fig. 5 in \citealt{bailes2020}).

In all cases the on-pulse spectra were well fit by single Gaussian function and in no case was the frequency response found to be highly asymmetric. Furthermore no GPs were found to exhibit band-limited emission across more than one fractional part of the band (e.g. two bright frequency knots) as would be more typical of interstellar scintillation. 

Band-limited emission has been observed in the interpulse (IP) of the Crab pulsar \citep{he07, Hankins2016}. However, this emission is mostly `banded', such that emission is observed in multiple regularly spaced frequency bands. The Crab main pulse (MP) in contrast does not show bandedness in general, but single band-limited emission, more closely resembling the flux knots of PSR J0540$-$6919, can be identified in some individual nanoshots within `sparse' MPs (e.g. Fig.~4 in \citealt{he07}).

Fig. \ref{fig:gallery}c to \ref{fig:gallery}e contain a subset of flux knots and their frequency responses. The narrowest (and second brightest) is shown in Fig.~\ref{fig:gallery}d. Analysing the giants discovered before and after this band-limited pulse we find that a broad-band GP (\gls{snr}$>15$) occurred just 35 rotations before from the same phase window. The first discovered GP after it occurred 541 rotations or 28\,s later.

In Fig.~\ref{fig:seq} we show a sequence of single pulses, where these frequency changes happen even faster: in just two rotations.  In this sequence we find four GPs in six rotations. The sequence of pulses are labeled \#1 to \#6, with rotations \#3 and \#5 not plotted since no radio emission was detected. 
The brightest broad-band pulse (\#4) is preceded by a bright band-limited pulse (\#2) just two rotations (or 100\,ms) before it. This is the fastest time-scale over which we see changes in the frequency behaviour of the detected GPs. We note that while pulses \#2 and \#4 are not emitted from the same phase window, the shortest time-scale over which we do observe emission bandwidth changes from the same phase window is 35 rotations (1.8\,sec), as described above and associated with Fig.~\ref{fig:gallery}d.  The mechanism responsible for the changes in the observed frequency emission (be it the interstellar medium or intrinsic to the GP emission) therefore has to be able to change on time-scales of just a few seconds.

The \gls{snr} > 100 band-limited pulse referred to in Sec.~\ref{sec:sg}, is shown in Fig.~\ref{fig:gallery}e and has emission predominantly in the lower part of the MeerKAT band. In the case of this bright GP, a pulse is observed when integrating the upper part of the frequency band above 1436\,MHz or 4$\sigma$ from its frequency peak. In other words, integrating above the dotted line in the dynamic spectrum of Fig.~\ref{fig:gallery}e produces the pulse shown in the top panel. This result suggests that for lower \gls{snr} flux knots a non-detection of emission outside the band-limited range, may well be due to a lack of sensitivity. Fig.~\ref{fig:gallery} shows the non-detections outside of the flux knot range for two more GPs when applying the same analysis. 

\subsubsection{Double giants}\label{sec:doub}

A low occurrence of double GPs was found for epochs 1 and 3, the most striking of which is shown in Fig.~\ref{fig:gallery}b. These are rotations in which GPs are simultaneously observed from both emitting phase windows. Double giants were first reported for the well-studied Crab pulsar in 2010 \citep{Karuppusamy2010}, and has since been observed for the globular cluster pulsar J1823-3021A \citep{Knight2007, Abbate2020}.

For each of the 865 GPs we computed the \gls{snr} in the alternate phase window (leading or trailing) to the one it was found in. The dynamic spectra of GPs with a \gls{snr} > 4.5 signal in the alternate phase window were then inspected for evidence of double GP emission. Epoch 1 delivered three convincing double GPs, whereas in epoch 2 no clear doubles were seen. In epoch 3, we found two double GPs, neither as bright as the epoch 1 double GPs. From the total of five double GPs detected, in only two cases (both from epoch 1) are the giants in each phase window bright enough that they would independently have been discovered with a \gls{snr} cut-off of 7. 

\subsection{Dispersion measure estimates}\label{sec:dm}
\begin{figure}
\centering
\includegraphics[width=\columnwidth]{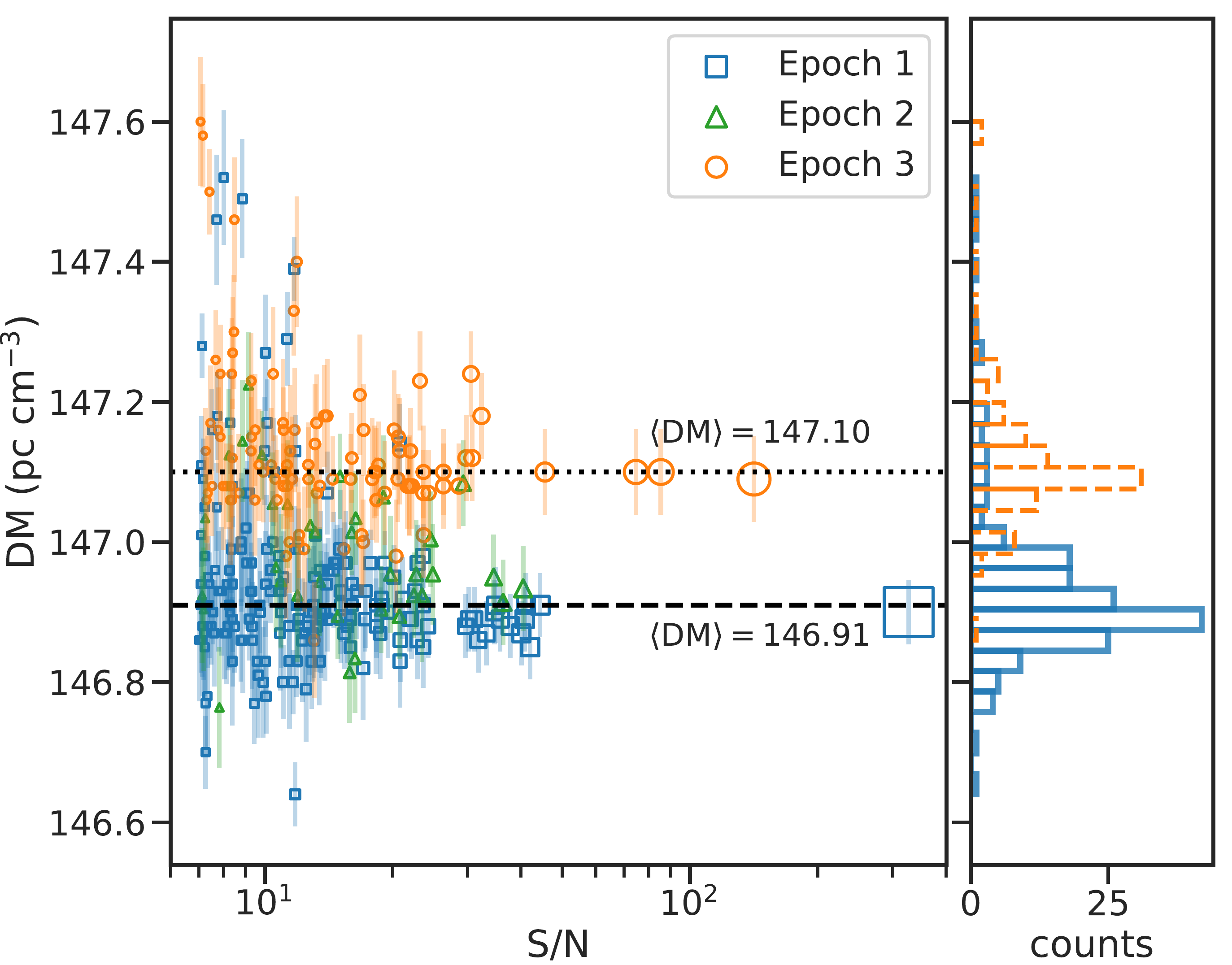}
\caption{Dispersion measure estimates for each GP, obtained by computing the DM values that maximise each GP's \gls{snr} across 32 frequency channels, and plotted per epoch. Marker sizes scale with \gls{snr} for better visibility of clustered data points. Only DM values for which error bars < 0.1\,pc\,cm$^{-3}$ are obtained are shown. The right hand panel shows the binned DM distributions for epochs 1 and 3. The median DM value changed from 146.91 to 147.10\,pc\,cm$^{-3}$ between these epochs.} \label{fig:dm}
\end{figure}

PSR J0540$-$6919 has the second highest DM of the 23 radio pulsars that have been discovered in the \gls{lmc}, with the highest value of 273\,pc\,cm$^{-3}$ measured for PSR J0537$-69$ \citep{Ridley2013}. All other \gls{lmc} pulsars have DM values ranging between 45\,pc\,cm$^{-3}$ and PSR J0540$-$6919's value of 147\,pc\,cm$^{-3}$. The NE2001 and YMW16 electron density models estimate a Galactic DM contribution of 54.98\,pc\,cm$^{-3}$ and 62.62\,pc\,cm$^{-3}$ along the line of sight to PSR J0540$-$6919 respectively \citep{Cordes2002,Yao2017}, i.e. less than half the pulsar's measured DM. As such the remaining 84--92\,pc\,cm$^{-3}$ is thought to be imparted mostly by the \gls{lmc}.

In Fig.~\ref{fig:dm} we show the DM distribution obtained when computing the \gls{snr} maximising DM value across 32 channels for each GP. The figure shows a clear distinction between the DM values obtained for the earlier epoch 1, and the DM values obtained five months later for epoch 3. The median value for each epoch is shown along the dashed and dotted lines.

We note that when we align the rising edges of the brightest GPs across 32 channels, or alternatively fit for scattering and DM corrections simultaneously as in \citet{Geyer2017}, we systematically obtain a lower DM value (by between 0.03 and 0.06\,pc\,cm$^{-3}$) than that of Fig.~\ref{fig:dm}. The motivation for these alternative DM computations can be seen from the averaged profiles of the brightest GPs (e.g. Fig.~\ref{fig:gallery}a) where the top of the pulse's rising edge appears skewed (after dedispersion to produce the highest \gls{snr}-valued GP). The rise-time of the GP is therefore not restricted to a single phase-bin. Such features could be indicative of the GPs encountering more than one scattering surface along the line of sight. 

\subsection{Pulse broadening}\label{sec:tau}
\begin{figure}
    \centering
    \includegraphics[width=0.9\columnwidth]{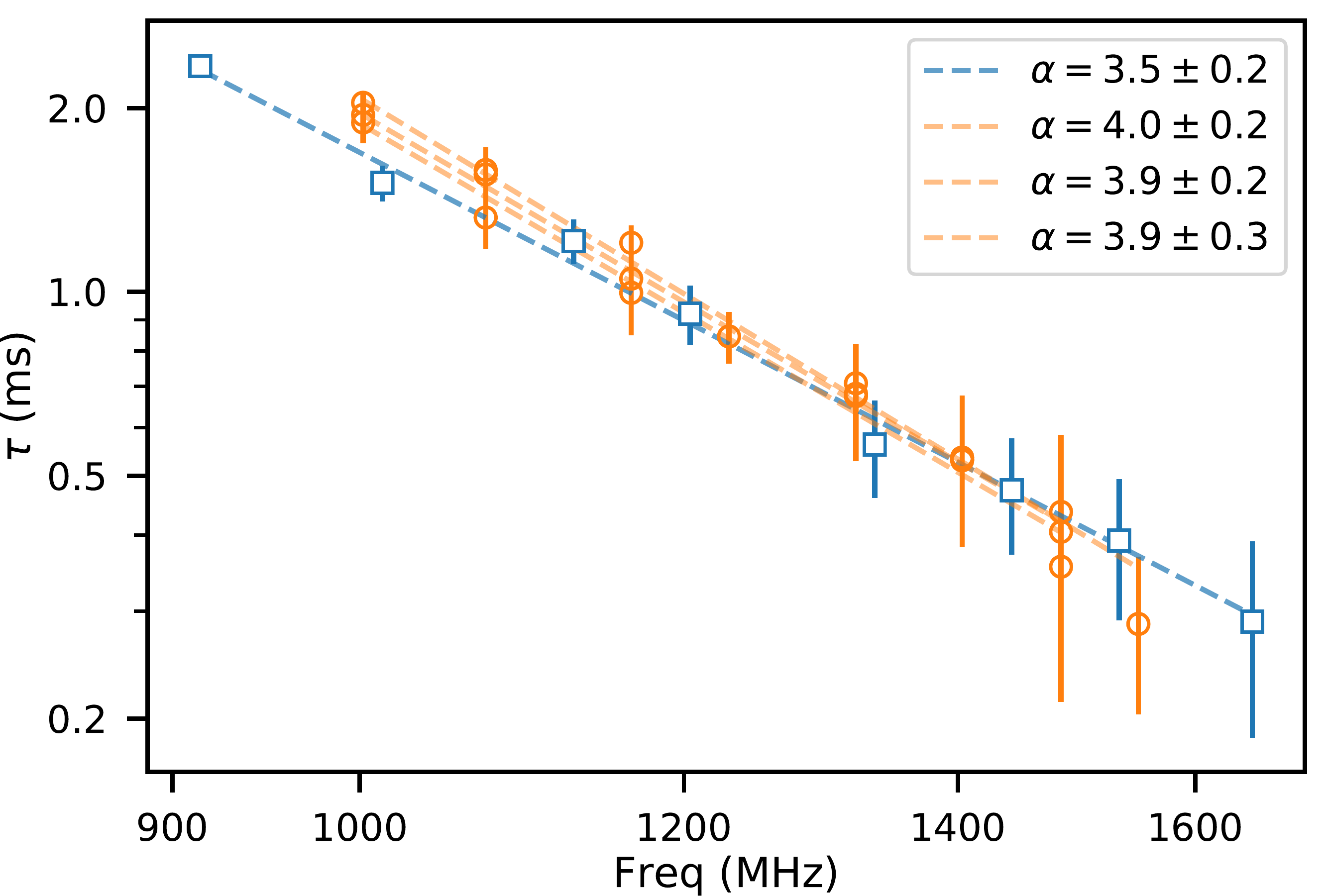}
    \caption{
  Plotted $\tau$ values represent GPs for which 6 or more out of 8 channels had an in-channel \gls{snr}>5.0. The power-law fits and their associated spectral indices ($\alpha$) are shown in dashed lines.  The $\tau$ errors have been increased uniformly across the band to the largest obtained error per GP, so that $\alpha$ values are not dominated by small error bars at the higher end of the spectrum where pulses are only weakly scattered. Similar $\alpha$ values are obtained, when not adjusting the errors, but instead fitting for $\tau$ values above 1500\,MHz. The $\tau$ values for brightest GP found (Fig \ref{fig:gallery}a, epoch 1) is shown with square makers. The other GPs in orange (circles) are from epoch 3.}\label{fig:scat}
\end{figure}

All the high \gls{snr} GP profiles that we detected have a clear exponential tail, typical of the multi-path propagation of radio emission through a dense and ionized interstellar medium. To model the scattering time-scales we used a single Gaussian pulse model convolved with the standard one-sided exponential ISM response ($e^{-t/\tau}$, where $\tau$ is the characteristic scattering time-scale, \citealt{Williamson1972}). The brightest GP (Fig.~\ref{fig:gallery}a) has a best fit $\tau$ value of  $0.92 \pm 0.02$\,ms at 1.2\,GHz and an unresolved intrinsic pulse width consistent with zero.

Obtaining an accurate $\tau$ distribution across the full GP sample is complicated by the large sample of lower \gls{snr} GPs, the covariances of $\tau$ with other Gaussian fitting parameters and the presence of band-limited pulses. The overall $\tau$ distributions resulting from scattering fits to frequency-averaged GP profiles, do not show evidence for correlated changes between $\tau$ and DM values.

To investigate the scattering time-scales more carefully, we consider GPs with \gls{snr} > 60, such that the data can be split into 8 channels and the frequency dependence of $\tau$ determined.  A subset of computed $\tau$ values and spectral indices ($\alpha$) are shown in Fig.~\ref{fig:scat}.  We account for low levels of scattering at the upper end of the MeerKAT band by increasing error bars uniformly. Spectral index values are mostly consistent with $\alpha = 4.0$ which describes isotropic Gaussian scattering by the ISM. We caution that obtained spectral index values are dependent on a particular ISM model, and the intrinsic pulse shape parameters \citep{Geyer2017}. The measured $\tau$ values across this brightest subset of GPs do show an increase in scattering time-scale from epoch~1 (blue, squares) to epoch 3 (orange, circles), and while not conclusive, could suggest a covariance with the observed changes in DM estimates.

\subsection{Rotation measure estimates}\label{sec:rm}
 
 Full Stokes data were recorded for epochs 2 and 3 and calibrated according to the procedure in Sec. \ref{sec:datared}. Using \texttt{rmfit} we compute the rotation measures (RM) for all GPs with \gls{snr} > 15 from these epochs. Of the 20 (epoch 2) and 44 (epoch 3) GPs above this threshold, RM estimates are obtained for 55 of them, but in only 9 cases are well converged RM estimates\footnote{using -r in \texttt{rmfit} to refine the RM estimation by splitting the band in two} found.  
 
 The most constraining single pulse RM estimate, $-246.5 \pm 0.3$\,rad\,m$^{-2}$, is obtained from the brightest GP across both epochs. Combining this with the second best constraint, we extend the RM error bars to provide a best fit RM value of $-245.8 \pm 1.0$\,rad\,m$^{-2}$. Both these estimates rely on epoch 3 data. The most constraining RM fit for a GP in epoch~2, observed four months earlier than epoch 3, is RM$=-230.5 \pm 2.0$\,rad\,m$^{-2}$; and the only other converging fit from this epoch lies within these error bars. If real, this would suggest a change in RM between epochs of \mbox{$15 \pm 2$\,rad\,m$^{-2}$}.

\subsection{Polarisation characteristics}

 \begin{figure}
    \centering
     \includegraphics[width=0.9\columnwidth]{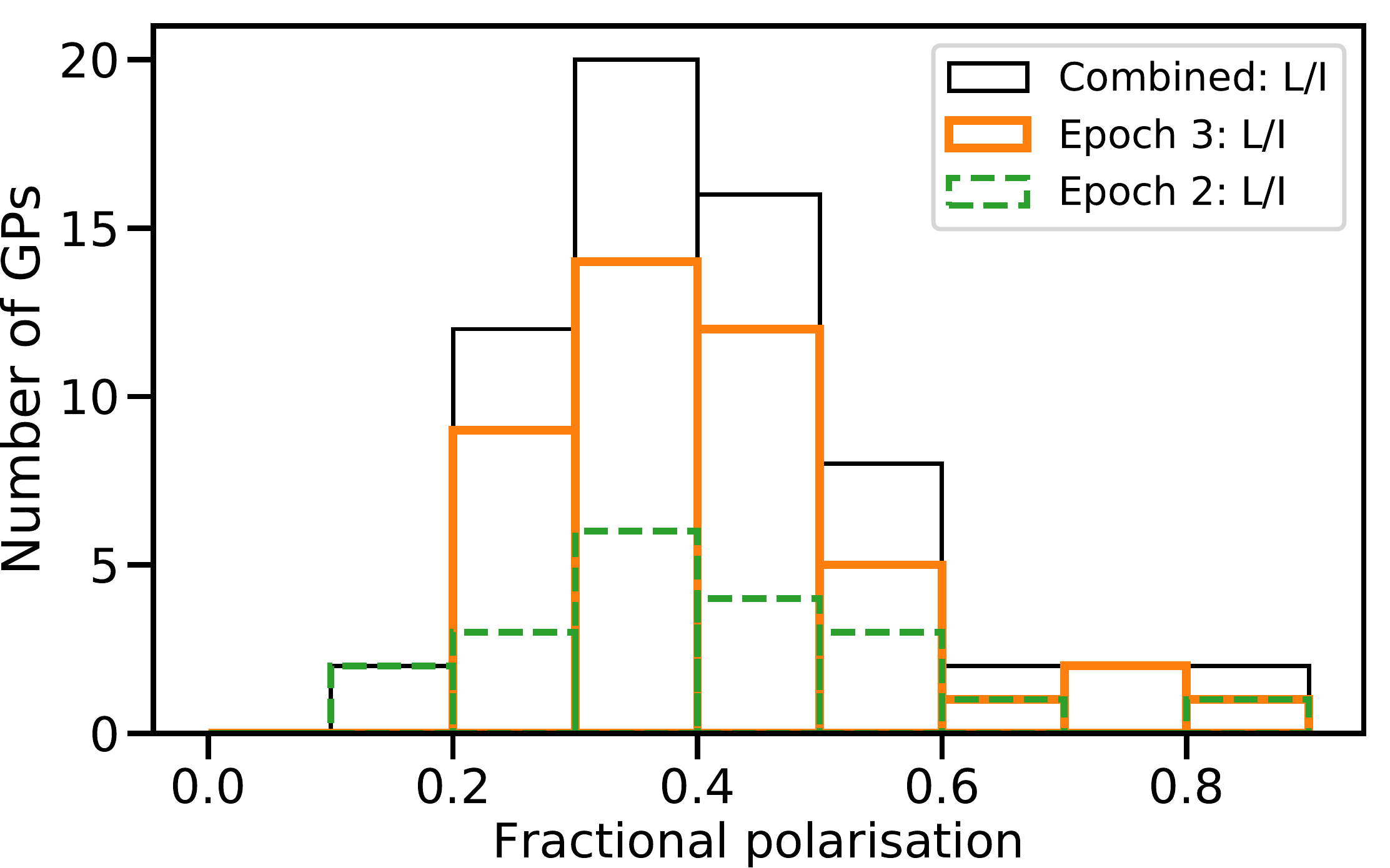}
    \caption{Fractional linear polarisation measurements obtained for GPs with \gls{snr} > 15 for epochs 2 and 3.  While some bright GPs do show linear polarisation fractions of 60\% and up, the majority of the high \gls{snr} GPs are found to have linear polarisation fractions of between 30 and 40\%.}
    \label{fig:linfrac}
\end{figure}
 
After computing the best-fit RM value, we corrected the RMs of all GPs accordingly. Adding together all the RM-corrected GPs in phase did not produce a measurable linear polarisation fraction for either epoch 2 or 3. 

\begin{figure}
    \centering
     \includegraphics[width=0.9\columnwidth]{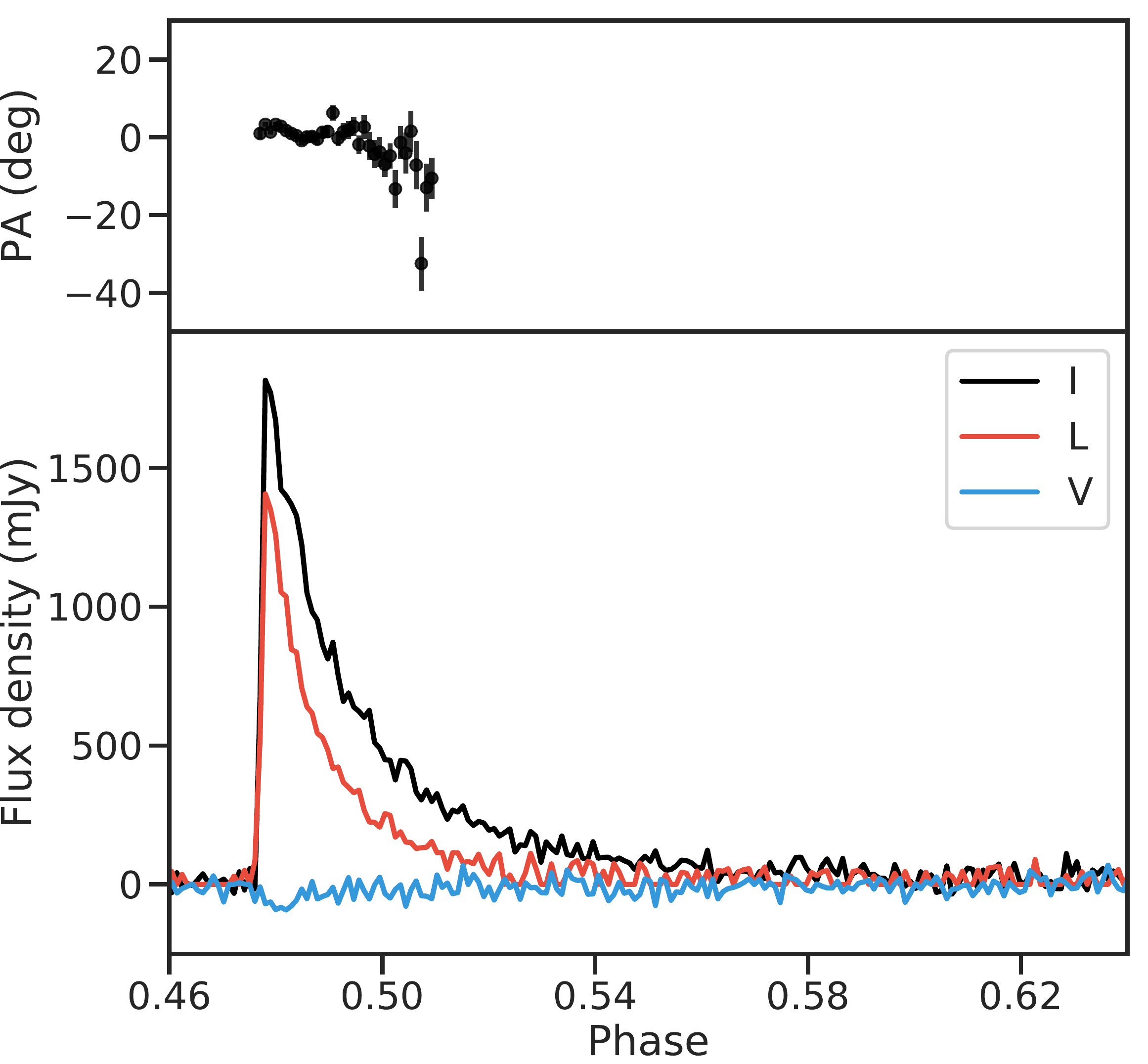}
      \includegraphics[width=0.9\columnwidth]{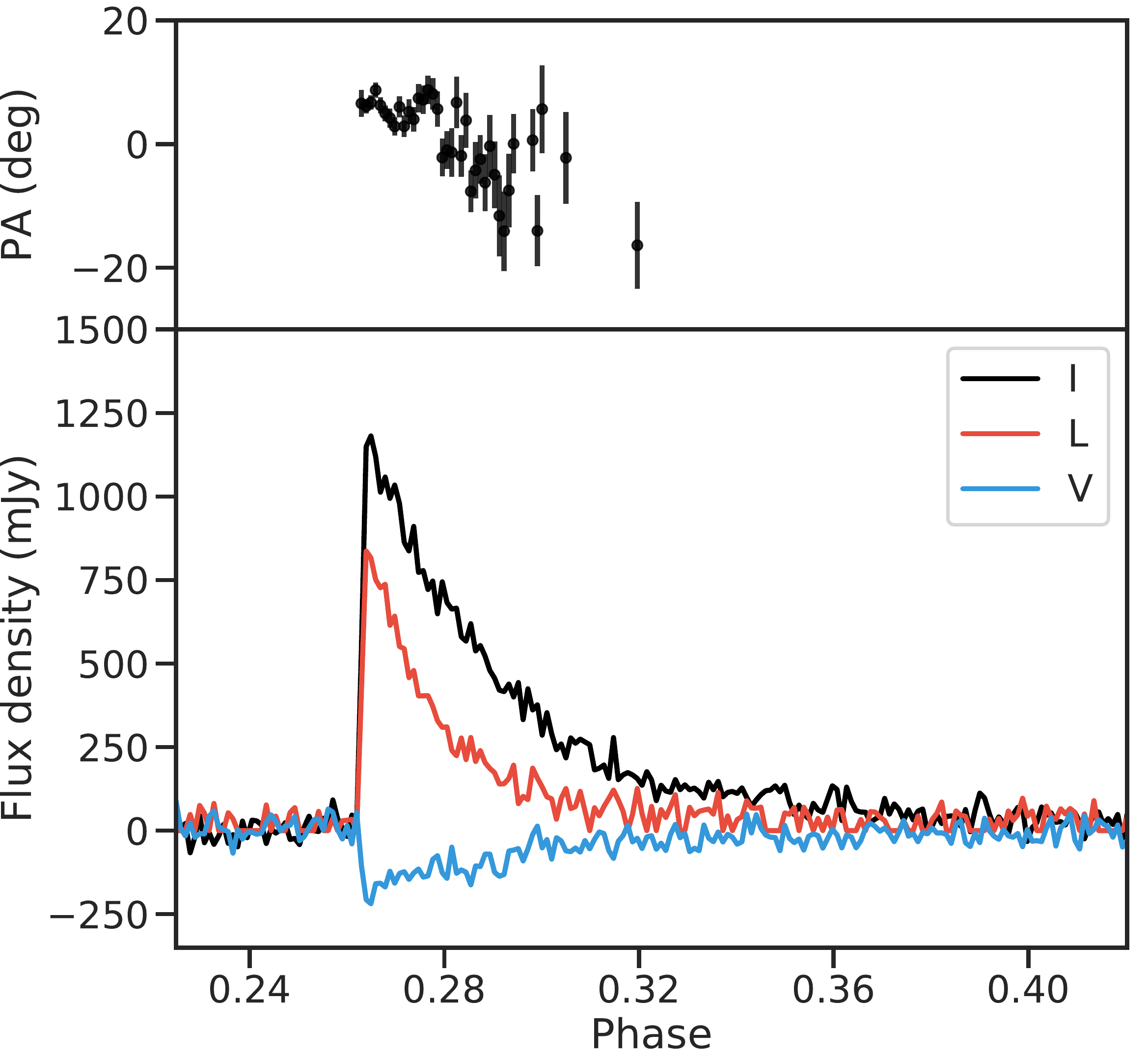}
    \caption{\textit{Top:} Polarisation profile of the (RM corrected) brightest broad-band GP of epoch 3. The profile is measured to have a linear polarisation fraction of 60.9\%. \textit{Bottom:} The RM corrected brightest band-limited GP of epoch 3 (also shown in Fig.~\ref{fig:gallery}e) has an estimated fractional linear polarisation of 51.5\%. All bright band-limited profiles are observed to shown significant linear fractional polarisation. In both plots the associated \gls{pa}s are shown for bins in which the linear polarisation (L) has a \gls{snr} > 4 with respect to the off-pulse \gls{rms}.}
    \label{fig:polprof}
\end{figure}

Fig.~\ref{fig:linfrac} shows the distribution in fractional linear polarisation obtained for GPs with \gls{snr} > 15. While some high \gls{snr} giants are linearly polarised with fractions above 50\%, the majority of these GPs are found to have linear polarisation fractions between 30 and 40\%, in closer agreement to the fractional value of 28\% obtained for the integrated profile in Fig. \ref{fig:ave}.  We do not obtain a reliable distribution of the lower \gls{snr} circular polarisation signals. However, we note that both positive and negative Stokes V signatures are observed with the maximum fractions between 20 and 30\% (e.g. Fig.~\ref{fig:polprof} bottom panel). 

In Fig.~\ref{fig:polprof} we show the calibrated, RM-corrected pulses of the brightest broad-band and the brightest band-limited GPs of epoch 3. The linear polarisation fractions are computed to be 60.9\% and 51.5\% respectively. The associated \gls{pa} is plotted for bins in which the linear polarisation has a \gls{snr} > 4 compared to the off-pulse \gls{rms} measured in Stokes I. We observe a swing in PA through the scattering tail of the GP in both cases.
 
Of the nine band-limited GPs with polarisation information seven have significant linear polarisation, as may well be expected if they are produced by narrow in frequency intrinsic coherent emission; and four have both measurable Stokes V and linear polarisation signatures. The two band-limited GPs without measurable linear polarisation are the weakest in the set (\gls{snr}<10), and it is likely due to a lack of sensitivity that the linear fraction is not well estimated.

\section{Discussion}\label{sec:disc}

\subsection{Flux density and power law comparisons to the literature}

The MeerKAT mean flux density values for PSR~J0540$-$6919 reported in this work are three to four times higher than that of Parkes, for which a value of 0.024\,mJy at 1.4 GHz was published \citep{jrmz04}. 
This discrepancy is somewhat larger than might normally be expected, but may in part be due to the differences in radio beam sizes of MeerKAT over Parkes, or due to refractive scintillation as discussed in Sec.~\ref{sec:refrac}. 

The power law index across the GP flux density distribution obtained in Sec.~\ref{sec:stat}, $-2.75 \pm 0.02$, is also much steeper than the published value of $-1.8 \pm 0.2$ in \citet{jrmz04}. The main reason for this is likely that \citet{jrmz04} fit a single power law across their obtained flux density values, which similar to our distribution show the onset of a flattening at low flux density values where the survey reaches its sensitivity limit. A single fit across all our data points provides a power law index of $-1.61 \pm 0.01$, but does not approximate the distribution well. Combined with the difference in observing sensitivity and statistics due to fewer GPs observed, a reasonably large discrepancy in power law indices might be expected.

In Fig.~\ref{fig:dist} we showed that while a power law fits the overall GP flux distribution well, we also observe local deviations or `kinks' in the distributions - the most prominent of which occur at a few mJy.  Interestingly, \citet{jrmz04} see the same trend (their Fig.~6), with the first kink appearing around 10\,mJy. This could ultimately point to the underlying flux distribution deviating from a simple power law.

The power law fit is also seen to deviate from the measured data points towards the higher end of the flux distribution. In studies with large GP samples a measurable flattening of the power law dependence at high GP flux densities has been observed; both for the Crab pulsar \citep{Mickaliger2012} and for PSR~B1937+21 \citep{McKee2019}. In contrast, analysing a sample of only the brightest Crab pulsar GPs (>130 Jy\,ms) \citet{Bera2019} do not report a flattening of the power law index, and measure a power law index $-2.81 \pm 0.05$ similar to what we compute for PSR J0540$-$6919, and in agreement with the value of $-2.79 \pm  0.01$ in \citet{Karuppusamy2010} for the Crab MP flux distribution.

For our study, with only a few data points at large flux values, we fail to constrain the power law index for $S_{\rm{GP}} > 15$\,mJy and note the supposed deviation is not yet significant. 

If the event rate estimates as obtained by the best-fit power law description holds at large flux values, we estimate that a GP at least as bright as our brightest detected GP will occur once in every 7 to 12 million rotations (or 20 to 30 observations equivalent to our data runs; Sec.~\ref{sec:fluxdistr}) with a 90\% probability. 

\subsection{Can all the observed emission be GP emission?}

Following the rationale in e.g. \cite{Knight2007} we can integrate eq.~\eqref{eq:prob} describing the power law of Fig.~\ref{fig:dist} to compute the lowest flux density values to which GP emission would have to extend to account for all of the detected flux. We find that if the GPs extended to flux density values of $0.024\times S_{\langle\rm{GP}\rangle}$ or 0.08\,mJy, then all detected flux can be fully described by a GP population. This value is in close agreement with the observed mean rotational flux density of 0.1\,mJy across the roughly 6~hr epoch. This strongly suggests that all the observed emission can be accounted for by GPs only. A GP-only distribution however, can only hold if the power-law describing the population continues unbroken for more than an order of magnitude to lower flux values. As mentioned previously, the kinks observed in the GP flux distributions could be evidence for a more complex underlying distribution.

We further note that many characteristics of the integrated profile does represent the GP aggregate. The averaged GP profile and the integrated profile are seen to overlap in profile shape, as shown in Fig.~\ref{fig:allgps}. The number of giants found per epoch also scales with the flux density measured for each epoch. The integrated pulse profile is shown to have a measurable, but low, linear polarisation fraction of 30\%, and while some high \gls{snr} GPs such as in Fig.~\ref{fig:polprof} are found to have linear polarisation fractions of 60\% or more, on the whole the GPs appear to have a low level of linear polarization (Fig.~\ref{fig:linfrac}). This is unusual for a pulsar with a high rate of loss in rotational energy, such as PSR J0540$-$6919 for which $\dot{E} = 10^{38.2}$\,erg\,s$^{-1}$. In \citet{Johnston2017} the authors show, from a sample of 600 pulsars observed by Parkes, that only five pulsars with $\dot{E} > 10^{35.5}$\,erg\,s$^{-1}$ have linear polarisation fractions less than 50\%. PSR J0540$-$6919 joins this small set of outliers in \citet{serylak2020arxiv}. The above similarities certainly suggest that a distribution of low \gls{snr} GPs, below our current single pulse detection, could make up the observed integrated profile.

\subsection{Is there evidence for GP clustering?}

In Sec.~\ref{sec:toa} we have shown that the overall GP time-of-arrival distribution is characteristic of GP emission independent of one another.
We now consider whether the double giants detected or the sequence of GPs in Fig.~\ref{fig:seq} suggests otherwise. Supplementary material, on how we compute the relevant probabilities presented here can be found in Appendix \ref{sec:app}.

We discovered five double GPs (Sec.~\ref{sec:doub}) two of which (both from epoch 1) have \gls{snr} > 7 for each pulse making up the double GP. Such simultaneous giant emission is rarely observed in other young giant emitting pulsars. 

In over 600 000 rotations of the Crab pulsar \citet{Karuppusamy2010} detected 197 double GPs with emission above 5$\sigma$ from both emitting phases. Analysing PSR B1937+21 data using the Large European Array for Pulsars (LEAP), \citet{McKee2019} reported no double giants, and estimated that a 95\% likelihood of observing a double GP would only be obtained after $\sim$~60 hrs of observing or 4.5 times their data set. Double GPs in PSR J1823-3021A were reported in \citet{Knight2007} and more recently in \citet{Abbate2020}. Using MeerKAT data, \citet{Abbate2020} found 74 double GPs (above their 7$\sigma$ threshold) in $\sim$ 3.3 million rotations. While \citet{Knight2007} suggested the detection of a double GP could be due to particular magnetospheric states, \citet{Abbate2020}, using a larger GP sample, concluded that the double GP detections are in line with Poisson statistics.

Assuming the emission from the trailing and leading phase windows is independent and the emission from one GP to the next is independent, as would hold for the Poisson description provided in Appendix \ref{sec:app}, we find that the probability of detecting at least two double GPs in epoch 1 is 0.17 (see Sec.~\ref{sec:double}). In other words we expect to detect two or more double GPs in one out of every 6 observations equivalent to epoch 1. As such we conclude that the double GPs detection rate does not suggest a deviation from the temporal Poissonian distribution. We note however that we do not detect any double giants from within the same phase window. Furthermore, as computed in Sec.~\ref{sec:double}, the above probability for detecting $\ge2$ double GPs decreases to one in 19 observations equivalent to epoch 1, if we associate independent probabilities with the leading and trailing emission windows. 

As presented in Fig.~\ref{fig:seq}, we also detected four GPs occurring in rapid succession during epoch 1. Using the Poisson description detailed in Sec.~\ref{sec:fourinsix} of Appendix \ref{sec:app}, we find that the detection of four or more GPs in an interval of 6 rotations would occur at least once in only every $\sim 7000$ observations equivalent to epoch 1. This probability decreases even further when taking into account from which emission window the GPs were observed. We find that the probability of detecting $\ge3$ trailing GPs and $\ge 1$ leading GPs in an interval of 6 rotations (as is descriptive of the sequence in Fig.~\ref{fig:seq}) at least once will occur in only 1 out of $\sim 30\times 10^3$ observations equivalent to epoch 1.

It therefore seems unlikely that this sequence of emission is coincidence, and could be valuable in modelling the GP emission mechanism as well as understanding the changes between broad-band and band-limited emission.

\subsection{Is there evidence for multiple scattering surfaces?}

As noted in Sec.~\ref{sec:dm}, the rise times of the GPs exceed our bin resolution. If the GPs are intrinsically narrow and well approximated by a delta function then these delays in rise times could be due to the emission encountering more than one scattering surface each imparting a frequency dependent shift on the peak's position. In such a two-screen model the pulse signal is convolved by a scattering function from each screen sequentially, leading to a rise time reflective of the shorter time-scale, and a scattering tail reflective of the longer time-scale. We measure the rise times across all high \gls{snr} GPs and find 2/512 bins or 0.2\,ms. For the highest \gls{snr} GP (Fig.\ref{fig:gallery}a) we confirm that the rise times scale steeply with frequency (across 8 channels) as would be expected for scattering, however we are too restricted by the phase bin resolution to resolve the rise times in the upper half of the band.

Multiple scattering screen models have been used to describe both the scattering observed in the Crab pulsar (e.g. \citealt{Rankin1973,Backer2000,Popov2017}), as well as the anomalous scattering measure for the Galactic Centre magnetar \citep{Bower2014,Desvignes2018}, and several FRBs \citep{Masui2015,Farah2018}. Multi-screen models are also often physically motivated by structures we expect the emission to cross along its line of sight. In the case of PSR~J0540$-$6919 we expect the dense surrounding nebula to impart the largest scattering time-scale on the radio pulses, while a secondary scattering time-scale would be due to its passage of the \gls{lmc} and the Milky Way.

\subsection{Are the flux knots caused by scintillation or lensing?}\label{sec:lens}

The scintillation decorrelation bandwidth obtained by $\Delta \nu_d = 1/(2\pi \tau)$, and using our scattering  time-scale ($\tau$) from Sec.~\ref{sec:tau}, is approximately 0.2\,kHz which is much smaller than 836\,kHz, the finest MeerKAT frequency resolution in its 1024 channel L-band mode. As such we are not sensitive to scintillation associated with the measured scattering time-scales, so that the much larger in frequency-extent flux knots can not be due to simple diffractive scintillation. 

Following arguments in \citet{Main2018}, we next investigate whether the band-limited nature of the flux knots could be caused via chromatic lens magnifications by the supernova shock boundary surrounding PSR J0540$-$6919. Such magnifications have been observed for bright single pulses of PSR~B1957+20. 

The shock boundary of the supernova remnant surrounding the central region powered by PSR J0540$-$6919 lies at a radial distance of 0.7\,pc from the pulsar \citep{mml+93}. If the flux knots we observe are lensed signals, they would persist for a characteristic time-scale associated with the lens resolution and the pulsar velocity. The flux knot's extent in frequency ($\Delta \nu$) would then provide an estimate of the lens magnification ($m$) through $\Delta \nu/\nu  \sim 1/m$, with $\nu$ the observing bandwidth. We estimate $\Delta \nu/\nu$ to be $\sim$10\% for the flux knot in Fig.~\ref{fig:gallery}d. The lens size $D$ is given by $D^2 \sim m \times \ell_{F}^2$, and the corresponding lens resolution $r$ is,

\begin{equation}
    r^2 \sim \frac{\ell_{F}^2}{m} \sim \frac{\Delta \nu}{\nu} \ell_{F}^2,
\end{equation}
with $\ell_{F} = \sqrt{\lambda\,d_{\rm{lens}}}$ the Fresnel scale. At the MeerKAT centre frequency ($\lambda = 23$\,cm) and taking the distance to the lens to be the shock boundary ($d_{\rm{lens}}=0.7$\,pc) the Fresnel scale is $\ell_{F} \approx 70\,000$\,km,  and consequently $r \approx 20\,000$\,km. 

Moving at a velocity of $\sim$ 100\,km/s the pulsar would therefore only cross this lens resolution element ($r$) in a few minutes, causing lensed signals to persist over such time-scales. Even when using the transverse velocity upper limit of 250\,km/s as published in \citet{Mignani2010}, the crossing time-scale remains more than a minute. 
The rapid changes observed at the few seconds level as shown in Sec.~\ref{sec:bl} can therefore not be described by this simple lensing model -- unless the plasma lens is much closer to the pulsar than the shock front  (< 1\,AU), which seems unlikely given the low density cavity expected from a supernova blast.

The scattering tails also appear to be unaffected by the presence of the knots (see Fig. \ref{fig:gallery}c and e); thus the GP emission either needs to be lensed near the pulsar before it is scattered, or the entire scattering screen would need to be amplified uniformly, leading to unphysically large magnifications.

We fail to describe the observed flux knots from PSR J0540$-$6919 through scintillation or lensing at the shock boundary, making it increasingly more likely that the band-limited nature of PSR J0540$-$6919 GPs are intrinsic to this young pulsar's emission mechanism. 

\subsection{Are epoch-to-epoch changes due to refractive scintillation?}\label{sec:refrac}

In Sec. \ref{sec:aveprof} we presented the mean flux density measurements for each epoch, and found the values to range from 0.07\,mJy to 0.13\,mJy (Table~\ref{tb:data}). As mentioned then, these longer term fluctuations could be due to refractive scintillation.

Having obtained the scattering time-scales in Sec.~\ref{sec:tau} we can now estimate the associated refractive scintillation time-scale. The refractive time-scale relates the time it takes the pulsar to traverse its associated scatter-broadened angular size across the scattering screen. The velocity with which this angular size-scale is crossed is the `scintillation velocity' ($v_{\rm{ISS}}$) expressed in terms of the velocity of the pulsar ($v_p$), the screen ($v_s$) and the earth ($v_E$) as,

\begin{equation}
v_{\rm{ISS}} = (1 - s)\,v_p - v_s + \,s\,\, v_E, 
\end{equation}

\noindent \citep{CordesLazio1998, Brisken2010} where $s$ defines the fractional distance of the scattering screen along the line of sight, $s = 1 - d_s / d_{\rm{psr}}$, with $d_{\rm{psr}} = 47.9$\,kpc the distance to PSR J0540$-$6919 and \mbox{$d_s = 47.9$\,kpc $-$ $0.7$\,pc} the distance to the scattering screen, both measured from the observer. As before, we estimate the scattering screen to be at the nebula shock boundary.  For this nebular screen $d_{\rm{psr}} \approx d_s$,  such that $s\approx 0$ and $v_{\rm{ISS}} \approx v_p - v_s$,  which is the relative velocity between the pulsar and the shock front. As before (in Sec.~\ref{sec:lens}) we assume this to be 100\,km/s.

Finally the refractive time-scale (with $s\approx 0$) is computed using,
\begin{equation}
t_{\rm{RISS}} = \frac{\pi \sqrt{2c\tau\,(d_{\rm{psr}} - d_{\rm{lens}})}}{v_p - v_s},
\end{equation}

\noindent which gives $t_{\rm{RISS}} \sim$ 40 days.

We therefore expect the refractive scintillation time-scale, associated with scattering in the nebula, to be weeks to months, which fits our data, and the findings in \citet{Johnston2004}.

\subsection{Nebular magnetic field estimates}
 
 Having obtained the first RM estimate towards PSR~J0540$-$6919 in Sec.~\ref{sec:rm}, we can compute the associated line of sight magnetic field.  Our best fit RM value, $-$245.8 rad\,m$^{-2}$, together with a DM measurement of 147.1\,pc\,cm$^{-3}$ leads to a magnetic field averaged along the line of sight of 2.06\,\si{\micro}G. This value is consistent with estimates of the \gls{lmc} magnetic field in \citet{Gaensler2005}, who found an ordered magnetic field component of $1.1$\si{\micro}G  and a random component of 3.6 times that, both to within a factor 2; as well as the upper limit on the total magnetic field of 7\,\si{\micro}G in \citet{Mao2012}.
 
We can also compute an estimate on the nebular magnetic field, if we assume that the estimated changes in RM and DM between epochs, obtained in Secs.~\ref{sec:dm} and \ref{sec:rm}, are dominated by changes within the dense medium of the supernova remnant.

Across five months, between March and August, we observed a DM variation of 0.16\,pc\,cm$^{-3}$.  In comparison, DM variations of 0.1\,pc\,cm$^{-3}$ have been measured over a five year time span \citep{McKee2018}, and variations of 0.03\,pc\,cm$^{-3}$ over 200 days \citep{Kuzmin2008} for the Crab pulsar. In the Crab nebula these changes are attributed to the relative motion between the pulsar and discrete high density structures of few AU (e.g. \citealt{McKee2018}). 

Combining our estimated changes in RM and DM we find, 

 \begin{equation}
 B_{\rm{{nebula}}} = 1.23\, \si{\micro}\rm{G}\times \Delta \rm{RM} / \Delta \rm{DM} \sim 115 \pm 15\, \si{\micro}G.
  \end{equation}
  
\noindent \citet{mml+93} found an estimate of 250\,\si{\micro}G based on the nebular age for PSR J0540-6919; and in studying the spectrum of the pulsar wind nebula across a broad range of frequencies \citet{Brantseg2013} presented nebular field strengths of $250 \pm 15$\,\si{\micro}G (using radio/optical/IR data) or $106^{+171}_{-68}$\,\si{\micro}G (using radio/X-ray data), the latter of which promotes a lower magnetic field in line with our computation. 

Building confidence in our estimates would require a longer term analysis of observed correlated changes in the DM and RM variations of PSR J0540$-$6919. In \citet{Rankin1988} such correlations were studied for the Crab pulsar leading to an estimate of the magnetic field between 150\,\si{\micro}G and 200\,\si{\micro}G. 

Overall, our GP estimated RM values are significantly less negative than what has been published from radio imaging data. Using 6~cm data from the Australian Telescope Compact Array, \citet{Brantseg2013} compute an average RM value of $-$346\,rad\,m$^{-2}$ across the pulsar wind nebula for SNR 0540$-$693, and higher negative value of $-$369\,rad\,m$^{-2}$ for the average RM across the supernova remnant shell. The authors however do measure strong variations in RM across the nebula, resulting in regions with both positive and negative RM measures, indicative of a clumpy medium, and which would complicate RM estimates. The regions for which they measure strong changes in RM handedness however lie to the western end of the shell and outside the MeerKAT beam.

\subsection{How do the observed flux knots compare with other GP emitting pulsars and FRBs?}

The MeerKAT sensitivity and bandwidth allowed us for the first time to distinguish between broad-band and highly band-limited GP emission from PSR~J0540$-$6919 -- raising exciting new questions about the emission mechanisms behind GPs and young neutron stars. 

The bright band-limited flux knots shown in this manuscript are the first to be published for any young pulsar other than the well-studied Crab pulsar. As described in Sec.~\ref{sec:bl}, the Crab IP showcases banded emission not observed in the MP which is mostly broadband. However, for some MPs where individual nanoshots can be distinguished the authors observe narrow in frequency emission \citep{he07, Hankins2016}. Since these differing pulse morphologies are seen when simultaneously observing IPs and MPs (i.e. over the same time-scales), the authors conclude that IP banded emission can not to be due to interstellar scintillation, but is instead intrinsic to the Crab's emission. \citet{Bilous2015} reported that the dynamic spectra of GPs from the energetic millisecond pulsar, PSR~B1821–24A, are also made up of multiple emission patches across the band. 

We do not find a single GP from PSR J0540$-$6919 with more than one flux knot across the band. While, given their similarities in ages and surroundings, we expect the GP emission mechanisms of the Crab and its `twin' to be related, there also appears to be clear observational differences.

On the other hand, there are striking observational similarities between the frequency structure of our observed \textit{flux knots}, and the band-limited emission observed from FRBs and especially repeating FRBs. The CHIME\footnote{Canadian Hydrogen Intensity Mapping Experiment}, ASKAP\footnote{Australian Square Kilometre Array Pathfinder } and UTMOST collaborations have all unveiled repeating FRB pulses that show complicated frequency structure and often highly band-limited emission \citep{Andersen2019,Fonseca2020,Kumar2019,Farah2018}.

In both \citet{Andersen2019} and \citet{Fonseca2020} many of the FRB repeating bursts are knot-like with typical emission bandwidths of 100-150\,MHz -- a similar frequency extent as the band-limited GPs we observe for PSR~J0540$-$6919. These FRB emissions are also equally symmetrical in frequency such that they are well-fitted by a single Gaussian function. The UTMOST burst had a particularly intense 2\,MHz knot of emission at 843 MHz.

Furthermore, the energies of the bursts from the original repeating FRB~121102 \citep{hss+19} have also been shown to follow a power law distribution \citep{lab+17}. Recent estimates of the power law index, with increasing numbers of bursts detected, are $-1.8 \pm 0.3$ \citep{Gourdji2019} and $-1.7 \pm 0.6$ \citep{Oostrum2020}. 

The association of FRB 121102 with a star forming region within its host galaxy as well as a continuum radio source suggestive of a pulsar wind nebula or supernova remnant \citep{Chatterjee2017,Bassa2017}, creates a strong connection between FRBs and young energetic pulsars and by extension makes a link to PSR J0540$-$6919 that resides in one of a handful of extra-galactic supernova remnants with associated radio pulsars observable to us.

A precise model of the emission mechanism remains up for debate. In the FRB literature, and especially in reference to repeating FRBs, band-limited emission models (often described as patchy emission) include both intrinsic cyclotron or synchrotron maser emission (e.g. \citealt{Metzger2019}), as well as burst propagation models relying on refractive lensing and diffractive scintillation by electron dense interstellar plasmas \citep{simard2020}. 

In the case of the PSR J0540$-$6919 flux knots we assert that it is intrinsic to the pulsar emission mechanism; 
or if not intrinsic, then caused in very close association with the pulsar, as the rapid time-scale of just a few pulse rotations over which we observe the frequency extent of the emission to change can not be explained by scintillation or plasma lensing at the shock front. 

\section{Conclusions}
\label{sec:conclusion}

The Southern Hemisphere MeerKAT telescope is providing a new and improved window on the Magellanic clouds including its radio pulsar population. In three relatively short observations of PSR J0540$-$6919, we recorded an increased detection rate of giant pulses and found a wealth of new information including a first estimate of the line of sight RM using this source and the pulsar's polarisation characteristics that exhibit low mean levels of linear polarisation. The MeerKAT bandwidth and sensitivity allowed a detailed investigation of the giant pulses' frequency structure unveiling band-limited flux knots occurring throughout the epochs. These provide a closer look at the intrinsic emission of one of the youngest and furthest pulsars known, and allows us to make tentative connections to the extra galactic Fast Radio Burst population.

\section{Acknowledgements} 
\label{sec:ack}
The MeerKAT telescope is operated by the South African Radio Astronomy
Observatory, which is a facility of the National Research Foundation,
an agency of the Department of Science and Innovation.
We thank Frank Marshall for an updated NICER X-ray ephemerides of PSR J0540$-$6919.
Parts of this research were conducted by the Australian Research Council Centre of 
Excellence for Gravitational Wave
Discovery (OzGrav), through project number CE170100004 and the project made
use of the OzSTAR supercomputer supported by AAL and ADACs through the gravitational
wave data centre.

\section{Data availability} 
The data underlying this article will be shared on reasonable request to the corresponding author.

\bibliographystyle{mnras}
\bibliography{0540}

\begin{thebibliography}{}
\makeatletter
\relax
\def\mn@urlcharsother{\let\do\@makeother \do\$\do\&\do\#\do\^\do\_\do\%\do\~}
\def\mn@doi{\begingroup\mn@urlcharsother \@ifnextchar [ {\mn@doi@}
  {\mn@doi@[]}}
\def\mn@doi@[#1]#2{\def\@tempa{#1}\ifx\@tempa\@empty \href
  {http://dx.doi.org/#2} {doi:#2}\else \href {http://dx.doi.org/#2} {#1}\fi
  \endgroup}
\def\mn@eprint#1#2{\mn@eprint@#1:#2::\@nil}
\def\mn@eprint@arXiv#1{\href {http://arxiv.org/abs/#1} {{\tt arXiv:#1}}}
\def\mn@eprint@dblp#1{\href {http://dblp.uni-trier.de/rec/bibtex/#1.xml}
  {dblp:#1}}
\def\mn@eprint@#1:#2:#3:#4\@nil{\def\@tempa {#1}\def\@tempb {#2}\def\@tempc
  {#3}\ifx \@tempc \@empty \let \@tempc \@tempb \let \@tempb \@tempa \fi \ifx
  \@tempb \@empty \def\@tempb {arXiv}\fi \@ifundefined
  {mn@eprint@\@tempb}{\@tempb:\@tempc}{\expandafter \expandafter \csname
  mn@eprint@\@tempb\endcsname \expandafter{\@tempc}}}

\bibitem[\protect\citeauthoryear{Abbate et~al.,}{Abbate
  et~al.}{2020}]{Abbate2020}
Abbate F.,  et~al., 2020, \mn@doi [\mnras] {10.1093/mnras/staa2510}, 498, 875

\bibitem[\protect\citeauthoryear{Andersen et~al.}{Andersen
  et~al.}{2019}]{Andersen2019}
Andersen B.,  et~al., 2019, \mn@doi [Astrophys. J. Lett.]
  {10.3847/2041-8213/ab4a80}, 885, L24

\bibitem[\protect\citeauthoryear{Backer, Wong  \& Valanju}{Backer
  et~al.}{2000}]{Backer2000}
Backer D.~C.,  Wong T.,   Valanju J.,  2000, \mn@doi [\apj] {10.1086/317150},
  543, 740

\bibitem[\protect\citeauthoryear{Bailes et~al.,}{Bailes
  et~al.}{2020}]{bailes2020}
Bailes M.,  et~al., 2020, Publ. Astron. Soc. Austral., 37, e028

\bibitem[\protect\citeauthoryear{Bassa et~al.,}{Bassa et~al.}{2017}]{Bassa2017}
Bassa C.~G.,  et~al., 2017, \mn@doi [\apj] {10.3847/2041-8213/aa7a0c}, 843, L8

\bibitem[\protect\citeauthoryear{{Bera} \& {Chengalur}}{{Bera} \&
  {Chengalur}}{2019}]{Bera2019}
{Bera} A.,  {Chengalur} J.~N.,  2019, \mn@doi [\mnras] {10.1093/mnrasl/slz140},
  \href {https://ui.adsabs.harvard.edu/abs/2019MNRAS.490L..12B} {490, L12}

\bibitem[\protect\citeauthoryear{Bilous, Pennucci, Demorest  \& Ransom}{Bilous
  et~al.}{2015}]{Bilous2015}
Bilous A.~V.,  Pennucci T.~T.,  Demorest P.,   Ransom S.~M.,  2015, \mn@doi
  [\apj] {10.1088/0004-637x/803/2/83}, 803, 83

\bibitem[\protect\citeauthoryear{Bower et~al.,}{Bower et~al.}{2014}]{Bower2014}
Bower G.~C.,  et~al., 2014, \apjl, 780, L2

\bibitem[\protect\citeauthoryear{Brantseg, McEntaffer, Bozzetto, Filipovic  \&
  Grieves}{Brantseg et~al.}{2013}]{Brantseg2013}
Brantseg T.,  McEntaffer R.~L.,  Bozzetto L.~M.,  Filipovic M.,   Grieves N.,
  2013, \mn@doi [\apj] {10.1088/0004-637x/780/1/50}, 780, 50

\bibitem[\protect\citeauthoryear{{Brisken}, {Macquart}, {Gao}, {Rickett},
  {Coles}, {Deller}, {Tingay}  \& {West}}{{Brisken} et~al.}{2010}]{Brisken2010}
{Brisken} W.~F.,  {Macquart} J.~P.,  {Gao} J.~J.,  {Rickett} B.~J.,  {Coles}
  W.~A.,  {Deller} A.~T.,  {Tingay} S.~J.,   {West} C.~J.,  2010, \mn@doi
  [\apj] {10.1088/0004-637X/708/1/232}, \href
  {https://ui.adsabs.harvard.edu/abs/2010ApJ...708..232B} {708, 232}

\bibitem[\protect\citeauthoryear{{Burke-Spolaor} et~al.,}{{Burke-Spolaor}
  et~al.}{2012}]{Burke-Spolaor2012}
{Burke-Spolaor} S.,  et~al., 2012, \mn@doi [\mnras]
  {10.1111/j.1365-2966.2012.20998.x}, \href
  {https://ui.adsabs.harvard.edu/abs/2012MNRAS.423.1351B} {423, 1351}

\bibitem[\protect\citeauthoryear{{Cairns}, {Johnston}, {Das}  \&
  {Robinson}}{{Cairns} et~al.}{2003}]{Cairns2003}
{Cairns} I.~H.,  {Johnston} S.,  {Das} P.,   {Robinson} P.~A.,  2003, in
  {Bailes} M.,  {Nice} D.~J.,   {Thorsett} S.~E.,  eds,  Astronomical Society
  of the Pacific Conference Series Vol. 302, Radio Pulsars. p.~191

\bibitem[\protect\citeauthoryear{Camilo et~al.,}{Camilo
  et~al.}{2018}]{Camilo2018}
Camilo F.,  et~al., 2018, \mn@doi [\apj] {10.3847/1538-4357/aab35a}, 856, 180

\bibitem[\protect\citeauthoryear{{Chatterjee} et~al.,}{{Chatterjee}
  et~al.}{2017}]{Chatterjee2017}
{Chatterjee} S.,  et~al., 2017, \mn@doi [\nat] {10.1038/nature20797}, \href
  {http://adsabs.harvard.edu/abs/2017Natur.541...58C} {541, 58}

\bibitem[\protect\citeauthoryear{{Cognard}, {Shrauner}, {Taylor}  \&
  {Thorsett}}{{Cognard} et~al.}{1996}]{cstt96}
{Cognard} I.,  {Shrauner} J.~A.,  {Taylor} J.~H.,   {Thorsett} S.~E.,  1996,
  ApJ, 457, L81

\bibitem[\protect\citeauthoryear{{Cordes} \& {Lazio}}{{Cordes} \&
  {Lazio}}{2002}]{Cordes2002}
{Cordes} J.~M.,  {Lazio} T.~J.~W.,  2002, arXiv: astro-ph/0207156, \href
  {http://adsabs.harvard.edu/abs/2002astro.ph..7156C} {}

\bibitem[\protect\citeauthoryear{{Cordes} \& {Rickett}}{{Cordes} \&
  {Rickett}}{1998}]{CordesLazio1998}
{Cordes} J.~M.,  {Rickett} B.~J.,  1998, \mn@doi [\apj] {10.1086/306358}, \href
  {https://ui.adsabs.harvard.edu/abs/1998ApJ...507..846C} {507, 846}

\bibitem[\protect\citeauthoryear{{Cordes} \& {Wasserman}}{{Cordes} \&
  {Wasserman}}{2016}]{cw16}
{Cordes} J.~M.,  {Wasserman} I.,  2016, MNRAS, 457, 232

\bibitem[\protect\citeauthoryear{Cordes, Bhat, Hankins, McLaughlin  \&
  Kern}{Cordes et~al.}{2004}]{Cordes2004}
Cordes J.~M.,  Bhat N. D.~R.,  Hankins T.~H.,  McLaughlin M.~A.,   Kern J.,
  2004, \mn@doi [\apj] {10.1086/422495}, 612, 375

\bibitem[\protect\citeauthoryear{{Desvignes} et~al.,}{{Desvignes}
  et~al.}{2018}]{Desvignes2018}
{Desvignes} G.,  et~al., 2018, \mn@doi [\apjl] {10.3847/2041-8213/aaa2f8},
  \href {https://ui.adsabs.harvard.edu/abs/2018ApJ...852L..12D} {852, L12}

\bibitem[\protect\citeauthoryear{Farah et~al.,}{Farah et~al.}{2018}]{Farah2018}
Farah W.,  et~al., 2018, \mn@doi [\mnras] {10.1093/mnras/sty1122}, 478, 1209

\bibitem[\protect\citeauthoryear{Fonseca et~al.,}{Fonseca
  et~al.}{2020}]{Fonseca2020}
Fonseca E.,  et~al., 2020, \mn@doi [\apj] {10.3847/2041-8213/ab7208}, 891, L6

\bibitem[\protect\citeauthoryear{{Foreman-Mackey}, {Hogg}, {Lang}  \&
  {Goodman}}{{Foreman-Mackey} et~al.}{2013}]{Foreman-Mackey2013}
{Foreman-Mackey} D.,  {Hogg} D.~W.,  {Lang} D.,   {Goodman} J.,  2013, \mn@doi
  [\pasp] {10.1086/670067}, \href
  {https://ui.adsabs.harvard.edu/abs/2013PASP..125..306F} {125, 306}

\bibitem[\protect\citeauthoryear{{G. Cusumano} et~al.,}{{G. Cusumano}
  et~al.}{2003}]{Cusumano2003}
{G. Cusumano} et~al., 2003, \mn@doi [A\&A] {10.1051/0004-6361:20031368}, 410,
  L9

\bibitem[\protect\citeauthoryear{Gaensler, Haverkorn, Staveley-Smith, Dickey,
  McClure-Griffiths, Dickel  \& Wolleben}{Gaensler et~al.}{2005}]{Gaensler2005}
Gaensler B.~M.,  Haverkorn M.,  Staveley-Smith L.,  Dickey J.~M.,
  McClure-Griffiths N.~M.,  Dickel J.~R.,   Wolleben M.,  2005, \mn@doi
  [Science] {10.1126/science.1108832}, 307, 1610

\bibitem[\protect\citeauthoryear{Geyer, Karastergiou, Kondratiev, Zagkouris,
  Kramer, Stappers  \& et al.}{Geyer et~al.}{2017}]{Geyer2017}
Geyer M.,  Karastergiou A.,  Kondratiev V.~I.,  Zagkouris K.,  Kramer M.,
  Stappers B.~W.,   et al. 2017, \mn@doi [MNRAS] {10.1093/mnras/stx1151}, 470,
  2659

\bibitem[\protect\citeauthoryear{Geyer, Buchner  \& Serylak}{Geyer
  et~al.}{2021}]{Geyer2021}
Geyer M.,  Buchner S.,   Serylak M.,  2021, \mn@doi
  [https://doi.org/10.48479/6ty4-jv90] {10.48479/6TY4-JV90}

\bibitem[\protect\citeauthoryear{Gourdji, Michilli, Spitler, Hessels, Seymour,
  Cordes  \& Chatterjee}{Gourdji et~al.}{2019}]{Gourdji2019}
Gourdji K.,  Michilli D.,  Spitler L.~G.,  Hessels J. W.~T.,  Seymour A.,
  Cordes J.~M.,   Chatterjee S.,  2019, \mn@doi [\apj]
  {10.3847/2041-8213/ab1f8a}, 877, L19

\bibitem[\protect\citeauthoryear{{Hankins} \& {Eilek}}{{Hankins} \&
  {Eilek}}{2007}]{he07}
{Hankins} T.~H.,  {Eilek} J.~A.,  2007, \mn@doi [ApJ] {10.1086/522362}, \href
  {http://adsabs.harvard.edu/abs/2007ApJ...670..693H} {670, 693}

\bibitem[\protect\citeauthoryear{{Hankins}, {Eilek}  \& {Jones}}{{Hankins}
  et~al.}{2016}]{Hankins2016}
{Hankins} T.~H.,  {Eilek} J.~A.,   {Jones} G.,  2016, \mn@doi [\apj]
  {10.3847/1538-4357/833/1/47}, \href
  {https://ui.adsabs.harvard.edu/abs/2016ApJ...833...47H} {833, 47}

\bibitem[\protect\citeauthoryear{Heiles, Campbell  \& Rankin}{Heiles
  et~al.}{1970}]{Heiles1970}
Heiles C.,  Campbell D.,   Rankin J.,  1970, \mn@doi [\nat] {10.1038/226529a0},
  226, 529

\bibitem[\protect\citeauthoryear{{Hessels} et~al.,}{{Hessels}
  et~al.}{2019}]{hss+19}
{Hessels} J.~W.~T.,  et~al., 2019, \apj, 876, L23

\bibitem[\protect\citeauthoryear{{Hotan}, {van Straten}  \&
  {Manchester}}{{Hotan} et~al.}{2004}]{psrchive}
{Hotan} A.~W.,  {van Straten} W.,   {Manchester} R.~N.,  2004, PASA, 21, 302

\bibitem[\protect\citeauthoryear{Jenet \& Gil}{Jenet \& Gil}{2003}]{Jenet2003}
Jenet F.~A.,  Gil J.,  2003, \mn@doi [\apj] {10.1086/379501}, 596, L215

\bibitem[\protect\citeauthoryear{Johnston \& Kerr}{Johnston \&
  Kerr}{2017}]{Johnston2017}
Johnston S.,  Kerr M.,  2017, \mn@doi [\mnras] {10.1093/mnras/stx3095}, 474,
  4629

\bibitem[\protect\citeauthoryear{{Johnston} \& {Romani}}{{Johnston} \&
  {Romani}}{2003}]{jr03}
{Johnston} S.,  {Romani} R.~W.,  2003, \apj, 590, L95

\bibitem[\protect\citeauthoryear{{Johnston} \& {Romani}}{{Johnston} \&
  {Romani}}{2004}]{Johnston2004}
{Johnston} S.,  {Romani} R.~W.,  2004, in {Camilo} F.,  {Gaensler} B.~M.,  eds,
   IAU Symposium Vol. 218, Young Neutron Stars and Their Environments. p.~315

\bibitem[\protect\citeauthoryear{Johnston, van Straten, Kramer  \&
  Bailes}{Johnston et~al.}{2001}]{Johnston2001}
Johnston S.,  van Straten W.,  Kramer M.,   Bailes M.,  2001, \mn@doi [\apj]
  {10.1086/319154}, 549, L101

\bibitem[\protect\citeauthoryear{{Johnston}, {Romani}, {Marshall}  \&
  {Zhang}}{{Johnston} et~al.}{2004}]{jrmz04}
{Johnston} S.,  {Romani} R.~W.,  {Marshall} F.~E.,   {Zhang} W.,  2004, MNRAS,
  355, 31

\bibitem[\protect\citeauthoryear{Johnston et~al.,}{Johnston
  et~al.}{2020}]{Johnston2020}
Johnston S.,  et~al., 2020, \mn@doi [\mnras] {10.1093/mnras/staa516}, 493, 3608

\bibitem[\protect\citeauthoryear{{Jonas} \& {MeerKAT Team}}{{Jonas} \& {MeerKAT
  Team}}{2016}]{Jonas2016}
{Jonas} J.,  {MeerKAT Team} 2016, in MeerKAT Science: On the Pathway to the
  SKA. p.~1

\bibitem[\protect\citeauthoryear{Kaplan, Chatterjee, Gaensler  \&
  Anderson}{Kaplan et~al.}{2008}]{Kaplan2008}
Kaplan D.~L.,  Chatterjee S.,  Gaensler B.~M.,   Anderson J.,  2008, \mn@doi
  [\apj] {10.1086/529026}, 677, 1201

\bibitem[\protect\citeauthoryear{{Karuppusamy}, {Stappers}  \& {van
  Straten}}{{Karuppusamy} et~al.}{2010}]{Karuppusamy2010}
{Karuppusamy} R.,  {Stappers} B.~W.,   {van Straten} W.,  2010, \mn@doi [\aap]
  {10.1051/0004-6361/200913729}, \href
  {https://ui.adsabs.harvard.edu/abs/2010A&A...515A..36K} {515, A36}

\bibitem[\protect\citeauthoryear{{Knight}}{{Knight}}{2007}]{Knight2007}
{Knight} H.~S.,  2007, \mn@doi [\mnras] {10.1111/j.1365-2966.2007.11810.x},
  \href {https://ui.adsabs.harvard.edu/abs/2007MNRAS.378..723K} {378, 723}

\bibitem[\protect\citeauthoryear{{Knight}, {Bailes}, {Manchester}, {Ord}  \&
  {Jacoby}}{{Knight} et~al.}{2006}]{kbm+06}
{Knight} H.~S.,  {Bailes} M.,  {Manchester} R.~N.,  {Ord} S.~M.,   {Jacoby}
  B.~A.,  2006, \apj, 640, 941

\bibitem[\protect\citeauthoryear{Kumar et~al.,}{Kumar et~al.}{2019}]{Kumar2019}
Kumar P.,  et~al., 2019, \mn@doi [\apj] {10.3847/2041-8213/ab5b08}, 887, L30

\bibitem[\protect\citeauthoryear{{Kuzmin, A.}, {Losovsky, B. Ya.}, {Jordan, C.
  A.}  \& {Smith, F. G.}}{{Kuzmin, A.} et~al.}{2008}]{Kuzmin2008}
{Kuzmin, A.} {Losovsky, B. Ya.} {Jordan, C. A.}  {Smith, F. G.} 2008, \mn@doi
  [A\&A] {10.1051/0004-6361:20079211}, 483, 13

\bibitem[\protect\citeauthoryear{{Law} et~al.,}{{Law} et~al.}{2017}]{lab+17}
{Law} C.~J.,  et~al., 2017, \apj, 850, 76

\bibitem[\protect\citeauthoryear{{Lorimer} \& {Kramer}}{{Lorimer} \&
  {Kramer}}{2004}]{pulsarhandbook}
{Lorimer} D.~R.,  {Kramer} M.,  2004, {\textit{Handbook of Pulsar Astronomy}}.
UK: Cambridge University Press

\bibitem[\protect\citeauthoryear{{Lorimer}, {Bailes}, {McLaughlin}, {Narkevic}
  \& {Crawford}}{{Lorimer} et~al.}{2007}]{lbm+07}
{Lorimer} D.~R.,  {Bailes} M.,  {McLaughlin} M.~A.,  {Narkevic} D.~J.,
  {Crawford} F.,  2007, Science, 318, 777

\bibitem[\protect\citeauthoryear{{Lundgren}, {Cordes}, {Ulmer}, {Matz},
  {Lomatch}, {Foster}  \& {Hankins}}{{Lundgren} et~al.}{1995}]{lcu+95}
{Lundgren} S.~C.,  {Cordes} J.~M.,  {Ulmer} M.,  {Matz} S.~M.,  {Lomatch} S.,
  {Foster} R.~S.,   {Hankins} T.,  1995, \apj, 453, 433

\bibitem[\protect\citeauthoryear{Main et~al.,}{Main et~al.}{2018}]{Main2018}
Main R.,  et~al., 2018, \mn@doi [Nature] {10.1038/s41586-018-0133-z}, 557, 522

\bibitem[\protect\citeauthoryear{{Manchester}, {Mar}, {Lyne}, {Kaspi}  \&
  {Johnston}}{{Manchester} et~al.}{1993}]{mml+93}
{Manchester} R.~N.,  {Mar} D.~P.,  {Lyne} A.~G.,  {Kaspi} V.~M.,   {Johnston}
  S.,  1993, \apj, 403, L29

\bibitem[\protect\citeauthoryear{{Mao} et~al.,}{{Mao} et~al.}{2012}]{Mao2012}
{Mao} S.~A.,  et~al., 2012, \mn@doi [\apj] {10.1088/0004-637X/759/1/25}, \href
  {https://ui.adsabs.harvard.edu/abs/2012ApJ...759...25M} {759, 25}

\bibitem[\protect\citeauthoryear{{Masui} et~al.,}{{Masui}
  et~al.}{2015}]{Masui2015}
{Masui} K.,  et~al., 2015, \mn@doi [\nat] {10.1038/nature15769}, \href
  {http://adsabs.harvard.edu/abs/2015Natur.528..523M} {528, 523}

\bibitem[\protect\citeauthoryear{McKee, Lyne, Stappers, Bassa  \& Jordan}{McKee
  et~al.}{2018}]{McKee2018}
McKee J.~W.,  Lyne A.~G.,  Stappers B.~W.,  Bassa C.~G.,   Jordan C.~A.,  2018,
  \mn@doi [\mnras] {10.1093/mnras/sty1727}, 479, 4216

\bibitem[\protect\citeauthoryear{{McKee} et~al.,}{{McKee}
  et~al.}{2019}]{McKee2019}
{McKee} J.~W.,  et~al., 2019, \mn@doi [\mnras] {10.1093/mnras/sty3058}, \href
  {https://ui.adsabs.harvard.edu/abs/2019MNRAS.483.4784M} {483, 4784}

\bibitem[\protect\citeauthoryear{Metzger, Margalit  \& Sironi}{Metzger
  et~al.}{2019}]{Metzger2019}
Metzger B.~D.,  Margalit B.,   Sironi L.,  2019, \mn@doi [\mnras]
  {10.1093/mnras/stz700}, 485, 4091

\bibitem[\protect\citeauthoryear{Mickaliger et~al.,}{Mickaliger
  et~al.}{2012}]{Mickaliger2012}
Mickaliger M.~B.,  et~al., 2012, \mn@doi [\apj] {10.1088/0004-637x/760/1/64},
  760, 64

\bibitem[\protect\citeauthoryear{{Middleditch} \& {Pennypacker}}{{Middleditch}
  \& {Pennypacker}}{1985}]{mp85}
{Middleditch} J.,  {Pennypacker} C.,  1985, Nature, 313, 659

\bibitem[\protect\citeauthoryear{{Mignani}, {Sartori}, {de Luca}, {Rudak},
  {S{\l}owikowska}, {Kanbach}  \& {Caraveo}}{{Mignani}
  et~al.}{2010}]{Mignani2010}
{Mignani} R.~P.,  {Sartori} A.,  {de Luca} A.,  {Rudak} B.,  {S{\l}owikowska}
  A.,  {Kanbach} G.,   {Caraveo} P.~A.,  2010, \mn@doi [\aap]
  {10.1051/0004-6361/200913870}, \href
  {https://ui.adsabs.harvard.edu/abs/2010A&A...515A.110M} {515, A110}

\bibitem[\protect\citeauthoryear{{Oostrum} et~al.,}{{Oostrum}
  et~al.}{2020}]{Oostrum2020}
{Oostrum} L.~C.,  et~al., 2020, \mn@doi [\aap] {10.1051/0004-6361/201937422},
  \href {https://ui.adsabs.harvard.edu/abs/2020A&A...635A..61O} {635, A61}

\bibitem[\protect\citeauthoryear{{Os{\l}owski}, {van Straten}, {Bailes},
  {Jameson}  \& {Hobbs}}{{Os{\l}owski} et~al.}{2014}]{ovb+14}
{Os{\l}owski} S.,  {van Straten} W.,  {Bailes} M.,  {Jameson} A.,   {Hobbs} G.,
   2014, \mn@doi [MNRAS] {10.1093/mnras/stu804}, \href
  {http://adsabs.harvard.edu/abs/2014MNRAS.441.3148O} {441, 3148}

\bibitem[\protect\citeauthoryear{Popov, Rudnitskii  \& Soglasnov}{Popov
  et~al.}{2017}]{Popov2017}
Popov M.,  Rudnitskii A.,   Soglasnov V.,  2017, \mn@doi [Astron. Rep.]
  {https://doi.org/10.1134/S1063772917030064}, 61, 178

\bibitem[\protect\citeauthoryear{{Rankin} \& {Counselman}}{{Rankin} \&
  {Counselman}}{1973}]{Rankin1973}
{Rankin} J.~M.,  {Counselman} C.~C. I.,  1973, \mn@doi [\apj] {10.1086/152099},
  \href {https://ui.adsabs.harvard.edu/abs/1973ApJ...181..875R} {181, 875}

\bibitem[\protect\citeauthoryear{Rankin, Campbell, Isaacman  \& Payne}{Rankin
  et~al.}{1988}]{Rankin1988}
Rankin J.~M.,  Campbell D.~B.,  Isaacman R.~B.,   Payne R.~R.,  1988, \aap,
  202, 166

\bibitem[\protect\citeauthoryear{Ridley, Crawford, Lorimer, Bailey, Madden,
  Anella  \& Chennamangalam}{Ridley et~al.}{2013}]{Ridley2013}
Ridley J.~P.,  Crawford F.,  Lorimer D.~R.,  Bailey S.~R.,  Madden J.~H.,
  Anella R.,   Chennamangalam J.,  2013, \mn@doi [\mnras]
  {10.1093/mnras/stt709}, 433, 138

\bibitem[\protect\citeauthoryear{{Romani} \& {Johnston}}{{Romani} \&
  {Johnston}}{2001}]{rj01}
{Romani} R.~W.,  {Johnston} S.,  2001, \apj, 557, L93

\bibitem[\protect\citeauthoryear{{Romani}, {Narayan}  \& {Blandford}}{{Romani}
  et~al.}{1986}]{rnb86}
{Romani} R.~W.,  {Narayan} R.,   {Blandford} R.,  1986, \mn@doi [MNRAS]
  {10.1093/mnras/220.1.19}, \href
  {https://ui.adsabs.harvard.edu/abs/1986MNRAS.220...19R} {220, 19}

\bibitem[\protect\citeauthoryear{Serylak et~al.,}{Serylak
  et~al.}{2020}]{serylak2020arxiv}
Serylak M.,  et~al., 2020, arXiv:astro-ph/2009.05797

\bibitem[\protect\citeauthoryear{{Seward}, {Harnden}  \& {Helfand}}{{Seward}
  et~al.}{1984}]{shh84}
{Seward} F.~D.,  {Harnden} F.~R. J.,   {Helfand} D.~J.,  1984, \apj, 287, L19

\bibitem[\protect\citeauthoryear{Simard \& Ravi}{Simard \&
  Ravi}{2020}]{simard2020}
Simard D.,  Ravi V.,  2020, \mn@doi [\apj] {10.3847/2041-8213/abaa40}, 899, L21

\bibitem[\protect\citeauthoryear{{Staelin} \& {Reifenstein}}{{Staelin} \&
  {Reifenstein}}{1968}]{sr68}
{Staelin} D.~H.,  {Reifenstein} Edward~C. I.,  1968, Science, 162, 1481

\bibitem[\protect\citeauthoryear{{Thornton} et~al.,}{{Thornton}
  et~al.}{2013}]{tsb+13}
{Thornton} D.,  et~al., 2013, Science, 341, 53

\bibitem[\protect\citeauthoryear{Walker}{Walker}{2011}]{Walker2011}
Walker A.~R.,  2011, \mn@doi [Astrophys. \& Space Science]
  {10.1007/s10509-011-0961-x}, 341, 43

\bibitem[\protect\citeauthoryear{{Williamson}}{{Williamson}}{1972}]{Williamson1972}
{Williamson} I.~P.,  1972, MNRAS, 157, 55

\bibitem[\protect\citeauthoryear{Yao, Manchester  \& Wang}{Yao
  et~al.}{2017}]{Yao2017}
Yao J.~M.,  Manchester R.~N.,   Wang N.,  2017, \mn@doi [\apj]
  {10.3847/1538-4357/835/1/29}, 835, 29

\bibitem[\protect\citeauthoryear{{Zhang}, {Marshall}, {Gotthelf}, {Middleditch}
   \& {Wang}}{{Zhang} et~al.}{2001}]{Zhang2001}
{Zhang} W.,  {Marshall} F.~E.,  {Gotthelf} E.~V.,  {Middleditch} J.,   {Wang}
  Q.~D.,  2001, \mn@doi [\apjl] {10.1086/321703}, \href
  {https://ui.adsabs.harvard.edu/abs/2001ApJ...554L.177Z} {554, L177}

\bibitem[\protect\citeauthoryear{{van Straten} \& {Bailes}}{{van Straten} \&
  {Bailes}}{2011}]{dspsr}
{van Straten} W.,  {Bailes} M.,  2011, PASA, 28, 1

\makeatother
\end{thebibliography}

\appendix
\section{The probability of detecting clustered giant pulses}\label{sec:app}

We are interested in calculating the probability of detecting at least $n$ giant pulses (GPs, events) in a sequence of $N$ consecutive pulses (an interval), given that each set of $N$ consecutive pulses is a sub-sample of the whole sample of $K$ pulses. The solution proceeds in two parts: in part A, we compute the probability of detecting $x \ge n$ events in an interval defined as $N$ consecutive pulses; in part B, we compute the probability of a super-event ($x \ge n$ events in $N$ pulses) in a set of $K/N$ independent intervals.  Under the assumption that all GPs are emitted independently, both parts satisfy the conditions of a Poisson point process, such that the distribution of the number of events in a given interval is described by the Poisson distribution, which has the cumulative distribution function,

\begin{equation}
F_n(k;\lambda\tau)=e^{-\lambda\tau} \sum_{j=0}^{k}{\frac {(\lambda\tau) ^{j}}{j!}}.
\end{equation}
Where $\lambda$ is the mean number of events per unit time and $\tau$ is the interval
of time under consideration, $F(k;\lambda\tau)$ is the probability
of detecting $n \le k$ events during this interval.

Furthermore, for a Poisson process, the distribution of times between successive events is exponentially distributed, such that

\begin{equation}
F_t(\tau;\lambda)=1-e^{-\lambda\tau}
\end{equation}
is the probability of detecting at least one event in the interval $0 \le t \le \tau$. Alternatively, for a given probability of detecting a single event, $P$, the above equation can be solved for the waiting time $\tau$, as
\begin{equation}
\tau = -\frac{\ln{(1-\rm{P})}}{\lambda}\label{eq:app3}.
\end{equation}

\subsection{Detecting four GPs in six rotations}\label{sec:fourinsix}

For part A, $\lambda_1 = n_\mathrm{obs} / K$ and $\tau_1 = N$, where $n_\mathrm{obs}$ is the total number of GPs observed in the whole sample of $K$ pulses and $N=6$ is the interval of interest. For epoch 1,
$n_\mathrm{obs}=441$ and $K=133188$; therefore, $\lambda_1\tau_1\simeq 0.019$ and the probability
of detecting $n \ge 4$ GPs in an interval of $N=6$ pulses is $p(n\ge4)=1-F_n(3;\lambda_1\tau_1)\simeq 6.4 \times 10^{-9}$.

For part B, note that $p(n\ge4)$ is the probability of success (detecting a super-event of $n \ge 4$ GPs in a single six-pulse interval) for a single trial (a six-pulse interval).  This probability is very small and the number of trials is large; therefore, the Poisson distribution with parameter 
$\lambda_2\tau_2 = p(n\ge4)$ K / N$ \simeq 1.4 \times 10^{-4}$ can be used to approximate the binomial distribution of the number of successes (detected super-events).

That is, the probability of detecting $n \ge 4$ GPs in an interval of $N=6$ pulses at least once in $K/N$ intervals is $p(x\ge1)=1-F_n(0;\lambda_2\tau_2)\simeq 1.4 \times 10^{-4}$.

Stated differently, the detection of 4 or more GPs in an interval of $N=6$ pulses would occur at least once in only 1 out of $\sim 7\times 10^3$ observations equivalent to epoch 1.

Under more careful consideration, we note that of the four GPs detected across six rotations (Fig. \ref{fig:seq}), three GPs are emitted from the trailing emission window and one from the leading emission window. Our above calculations can therefore be adapted to include independent event rates associated with the leading and trailing phase windows respectively. 

The leading emission window event rate ($\lambda_\mathrm{L} =  n_\mathrm{obs_L} / K$) and trailing emission window event rate ($\lambda_\mathrm{T} =  n_\mathrm{obs_T}/ K $), are evaluated using $K=133188$ as before with $n_\mathrm{obs_L} =229$ and $n_\mathrm{obs_T} =212$ the number of leading and trailing GPs in epoch 1 (cf. Table\ref{tb:giants}).

The probability of detecting at least three \textit{trailing} GPs in six rotations, ($\tau_1 = N = 6$) is given by, $p(n_T\ge3)=1-F(2;\lambda_T \tau_1)\simeq 1.4\times10^{-7}$. Similarly the probability of detecting at least one \textit{leading} GP in six rotations is given by, $p(n_L\ge1)=1-F(0;\lambda_L \tau_1)\simeq 0.0103$; such that the combined probability of detecting at least one leading GP and at least three trailing GPs in six rotations is, $p(n_T\ge3;n_L\ge1) \simeq1.4\times10^{-7} \times 0.0103\simeq 1.5 \times10^{-9}$.

Part B of the calculation proceeds as before, with $\bar{\lambda_2}\, \tau_2 = p(n_T\ge3;n_L\ge1) \, (K/N) \simeq 3.3 \times 10^{-5}$. It follows that the probability of detecting $n_T\ge3$ trailing GPs and $n_L\ge 1$ leading GPs in an interval of $N=6$ pulses \textit{at least once} in $K/N$ trails is,  $p(x\ge1)=1-F(0;\bar{\lambda_2}\tau_2)\simeq 3.3 \times 10^{-5}$, or once in only 1 out of $\sim 30\times 10^3$ observations equivalent to epoch 1.

\subsection{Detecting double GPs}\label{sec:double}

The same argument can be applied to compute the probability of detecting
double GPs.  The probability of detecting $n \ge 2$ GPs in an interval spanning one pulse is $p(n\ge2)=1-F_n(1;\lambda_1\tau_1^\prime)\simeq 5.5 \times 10^{-6}$, with $\tau_1^\prime=N=1$. 

For part B, $\tau_2^\prime=K/N=133188$, $\lambda_2^\prime = p(n\ge2)$,
$\lambda_2^\prime\tau_2^\prime \simeq 0.73$ and the probability of detecting $n \ge 2$ GPs in a single-pulse interval at least two times (as observed for epoch 1) in $K=133188$ pulses is $p(x\ge2)=1-F_n(1;\lambda_2^\prime \tau_2^\prime) \simeq 0.17$.  That is, the detection of 2 or more double GPs (where double GPs here also include combinations of GPs made up of more than doubles, e.g. tripple GPs) would occur in 1 out of $\sim 6$ observations equivalent to epoch 1.

As for the calculations in Sec.~\ref{sec:fourinsix}, the above can be adapted to take the independent event rates of the two emission windows into account. The associated event rates, 
$\lambda_L$ and $\lambda_T$, are as in Sec.~\ref{sec:fourinsix}, and $\tau_1\prime =1 $.

The probability of detecting at least one GP from the leading window is given by, 
$p(n_L\ge1)=1-F(0;\lambda_L \tau_1^\prime)\simeq 1.7\times 10^{-3}$ and similarly for the trailing window, $p(n_T\ge1)=1-F(0;\lambda_T \tau_1^\prime)\simeq 1.6\times 10^{-3}$. The probability of detecting at least one GP from each emission window simultaneously is then, 
$p(n_T\ge1;n_L\ge1) \simeq1.7\times10^{-3} \times 1.6\times 10^{-3} \simeq 2.7 \times 10^{-6}$.

For part B, we compute the probability of detecting at least one leading GP and one trailing GP in a single rotation at least two times in $K=133188$ pulses. We find $\lambda_2^{\prime\prime} \tau_2^\prime = p(n_T\ge1;n_L\ge1) \times K \simeq 0.36$, where $\tau_2^\prime=K$, such that $p(x\ge2)=1-F_n(1;\lambda_2^{\prime\prime} \tau_2^\prime) \simeq 0.052$. That is, the detection of two or more double-giant pulses (or higher multiples of GPs), taking into account the independent probabilities of the two emission windows, would occur in 1 out of $\sim 19$ observations equivalent to epoch 1.
\end{document}